\documentclass[iop,12pt]{emulateapj}
\usepackage{apjfonts}
\usepackage{epsfig}
\usepackage{natbib}
\usepackage{amsfonts}
\usepackage{amsmath}
\usepackage{multirow}
\usepackage{enumerate}
\usepackage[usenames]{color}
\newcommand\cqg{{Class. Quantum Grav.}}

\def\msun{M$_\odot$}

\def\Dwa{$\,$\uppercase\expandafter{\romannumeral5}$\,$}

\def\sless{\lower2pt\hbox{$\buildrel {\scriptstyle <}
   \over {\scriptstyle\sim}$}}

\def\sgreat{\lower2pt\hbox{$\buildrel {\scriptstyle >}
   \over {\scriptstyle\sim}$}}
\def\sharpnull#1{}

\def\isoni{$^{56}{\rm Ni}$}

\def\gray{$\gamma$-ray}
\def\grays{$\gamma$-rays}
\def\pns{proto-neutron star}

\def\bh{black hole}
\def\isco{innermost stable circular orbit}

\newcommand{\code}[1]{\texttt{#1}}

\begin{document}
%-----------------------------------------------------
\slugcomment{Accepted for publication in The Astrophysical Journal, May 18, 2012}
%\slugcomment{Draft version \today}

\title{The Arduous Journey to Black-Hole Formation in Potential Gamma-Ray Burst Progenitors}

\author{Luc Dessart,\altaffilmark{1,2} Evan O'Connor,\altaffilmark{2} and Christian
  D. Ott\altaffilmark{2,3,*}}
   \altaffiltext{1}{Laboratoire d'Astrophysique de Marseille, Universit\'e Aix-Marseille \& CNRS,
UMR7326, 38 rue Fr\'ed\'eric Joliot-Curie, 13388 Marseille, France, Luc.Dessart@oamp.fr }
  \altaffiltext{2}{TAPIR, Mailcode 350-17,
  California Institute of Technology, Pasadena, CA 91125, USA,
  evanoc@tapir.caltech.edu, cott@tapir.caltech.edu}
\altaffiltext{3}{Institute for the Physics and
 Mathematics of the Universe (IPMU), The University of Tokyo, Kashiwa, Japan}
\altaffiltext{*}{Alfred P. Sloan Research Fellow}

\begin{abstract}
  We present a quantitative study on the properties at death of
  fast-rotating massive stars evolved at low-metallicity, objects that
  are proposed as likely progenitors of long-duration \gray\ bursts
  (LGRBs).  We perform 1D+rotation stellar-collapse simulations on the
  progenitor models of \citet{woosley:06} and critically assess their
  potential for the formation of a black hole and a Keplerian disk
  (namely a collapsar) or a proto-magnetar.  We note that theoretical
  uncertainties in the treatment of magnetic fields and the
  approximate handling of rotation compromises the accuracy of
  stellar-evolution models.  We find that only the fastest rotating
  progenitors achieve sufficient compactness for black-hole formation
  while the bulk of models possess a core density structure typical of
  garden-variety core-collapse supernova (SN) progenitors evolved
  without rotation and at solar metallicity.  Of the models that do
  have sufficient compactness for black-hole formation, most of them
  also retain a large amount of angular momentum in the core, making
  them prone to a magneto-rotational explosion, therefore
  preferentially leaving behind a proto-magnetar.  A large progenitor
  angular-momentum budget is often the sole criterion invoked in the
  community today to assess the suitability for producing a
  collapsar. This simplification ignores equally important
  considerations such as the core compactness, which conditions
  black-hole formation, the core angular momentum, which may foster a
  magneto-rotational explosion preventing black-hole formation, or the
  metallicity and the residual envelope mass which must be compatible
  with inferences from observed LGRB/SNe.  Our study suggests that
  black-hole formation is non trivial, that there is room for
  accommodating both collapsars and proto-magnetars as LGRB
  progenitors, although proto-magnetars seem much more easily produced
  by current stellar-evolutionary models.
\end{abstract}


\section{Introduction/Context}

All stars with masses initially between $\sim$8 and
$\sim$150\,\msun\ eventually form a degenerate core that inevitably
collapses to form a \pns.  Much less certain is its subsequent
evolution, the potential formation of a \bh, and the powering of a
supernova (SN) explosion, sometimes associated with a long-duration
\gray\ burst (LGRB).  The situation is deceptively simple and the
outcome rests fundamentally on the solution to an energy problem.  An
explosion or a fizzle depends on the efficiency with which the system
can extract the prodigious gravitational energy released during
collapse. There are two principal forms of energy at disposal. The
first one is the gravitational-binding energy liberated by the
collapsing star and in particular its degenerate core.  The second is
its rotational energy (actually drawn from gravitational energy),
which is a function of the angular-momentum distribution and budget in
the progenitor star.  Understanding how these two energy sources can
be channeled to power relativistic and non-relativistic ejecta in
core-collapse SNe and leave behind a neutron star, a fast-rotating
pulsar, a magnetar, or a black hole, has been the subject of numerous
studies and the source of much debate \citep{bethe:90, woosley:93,
  herant:94, bhf:95, jankamueller:96, wheeler:00, kitaura:06,
  buras:06a, burrows:06, murphy:08, nordhaus:10, pejcha:11,hanke:11,
  takiwaki:11}.

Thermal MeV neutrinos are abundantly radiated from the
optically-thick, dense, and hot \pns, allowing its internal energy to
be released on a diffusion timescale. In the neutrino mechanism for
core-collapse SN explosions \citep{bethewilson:85}, the absorption in
the infalling mantle of $\sim$10\% of this neutrino flux may alone
lead to the revival of the stalled shock and the ejection of the
progenitor envelope with an asymptotic kinetic energy of up to 1\,B
($10^{51}\,\mathrm{erg}$) \citep{kitaura:06,buras:06b,buras:06a}.  But
this generic mechanism should not, however, be the origin of the
larger explosion energies of $\sim$\,10\,B inferred for a small
fraction of core-collapse SNe.  Instead, their scarcity calls for
exceptional circumstances, which seem intricately related to fast
progenitor-core rotation \citep{burrows:07b,takiwaki:11}.  It is
probable that most stars contain some angular momentum at the time of
death, either because they did not lose it completely through the
combined effects of magnetic torques and stellar-wind mass loss
\citep{maeder:00, meynet:00, heger:00, hirschi:04, meynet:05,
  hirschi:05, heger:05}, or because they gained it from a companion
star in a binary system \citep{wellstein:99,petrovic:05,cantiello:07}.
As the envelope collapses, the rotational energy increases.
During this process, the inner core
($\lesssim\,0.5\,M_\odot$) spins up by about three orders of magnitude and
remains in solid body rotation, while the outer core develops a
differentially-rotating profile \citep{ott:06spin}.  The energy
associated with rotation can be large ($\mathcal{O}(10\,\text{B})$)
and tapped by instabilities developing at the surface of the \pns, in
particular the magneto-rotational instability (MRI;
\citealt{balbus:91,akiyama:03,obergaulinger:09}).  Numerical
simulations for fast-rotating progenitor stars suggest that the
magneto-rotational mechanism of explosion is promising and offers a
very attractive explanation for the existence of highly-energetic SNe
\citep{leblanc:70, bisno:76, wheeler:00, yamada:04, moiseenko:06,
  burrows:07b, dessart:08a, takiwaki:11}. However, this mechanism
relies fundamentally on the assumption that the MRI can increase the
magnetic field to the required values. An assumption that has not yet
been shown numerically in the full core-collapse context, although
preliminary investigations in this sector are promising
\citep{obergaulinger:09}.   Hence, combined with the diversity of
progenitor-core properties, these two mechanisms alone, the neutrino
and the magneto-rotational mechanism, may explain the diversity of
core-collapse SNe, potentially encompassing two orders of magnitude in
explosion energy, from the low-luminosity SNe II-Plateau
\citep{pastorello:04} to highly energetic SNe Ic \citep{mazzali:02}.

A great puzzle is then to understand the necessary departures from
this general core collapse scenario to produce an LGRB in addition to
a SN explosion, as spectroscopically confirmed in, to date, six
LGRB/SNe pairs, (for a recent compilation, see \citealt{berger:11}).
This requires that $\sim$\,0.1-1\,B be injected into a low-mass,
baryon-deficient collimated region (at the origin of \grays) and that
at the same time $\sim$10\,B be injected quasi-isotropically to eject
the progenitor envelope (at the origin of the SN thermal emission
observed in the optical).  The very low occurrence rate of LGRB/SN per
core-collapse SN of $\lesssim$\,1\% \citep{guetta:07} calls for
progenitor properties that are rarely encountered in star
formation/evolution.  Interestingly, an unambiguous diversity emerges
among LGRB/SN observations, necessarily translating into a significant
range for the inferred properties of the SN ejecta,
with proposed masses and kinetic energies possibly varying by a factor
of 5--10 for both \citep{berger:11}.  Unfortunately, a significant
uncertainty is associated with such inferences. For example,
\citet{iwamoto:98} propose an ejecta mass of 11\,\msun\ with a total
energy of 20--50\,B for GRB980425/SN1998bw, but \citet{woosley:99}
reproduce the light curve with an ejecta of $\sim$5\,\msun\ and a
  total energy of 22\,B. Such differences are not surprising since
both spectra and light curves must be modeled simultaneously and with
allowance for the complicated non-LTE non-thermal and time-dependent
effects controlling the radiative transfer. The exceptionally fast
ejecta expansion of hypernovae is expected to strengthen the
time-dependent effects seen in ``standard" core-collapse SNe
\citep{dessart:08} while the large production of \isoni\ and
significant mixing may sizably affect line-profile shapes from which
the expansion rate is inferred \citep{dessart:12}.

Two LGRB central-engine models are currently favored.  They suggest
the key components for a successful LGRB/SN are a compact progenitor
with a short light-crossing time of $\sim$\,1\,s and fast rotation at
the time of collapse. One is the collapsar model \citep{woosley:93,
  macfadyen:99, macfadyen:01}: A fast-rotating progenitor fails to
explode in its early post-bounce phase and instead forms a black hole,
while the in-falling envelope eventually forms a Keplerian disk
feeding the hole on an accretion/viscous timescale comparable to that
of the LGRB. It is within this disk that the SN explosion is triggered
and the \isoni\ synthesized.  The other model involves a
proto-magnetar \citep{wheeler:00, bucciantini:08, metzger:10c,
  metzger:11} in which the LGRB is born after a successful SN
explosion (either by the neutrino or the magneto-rotational mechanism,
although the latter seems more likely given the rapid
rotation required for the magnetar) and the ejection of the overlying
envelope (or at least the onset of the ejection of the inner envelope
layers that clear the \pns\ surface). Fast rotation in the
\pns\ permits the huge enhancement of the magnetic-field energy and
stresses, which strengthen as the \pns\ cools and contracts,
eventually giving rise to relativistic ejecta ($\sim$\,10\,s after the
onset of collapse, once the neutrino driven wind decays away).  In
both models, rotation is key to control the dynamics in a unique
way. It is also key to allow the simultaneous ejection of
baryon-deficient material at relativistic speeds over a small solid
angle and the quasi-spherical ejection of the progenitor envelope.

A critical difference between the collapsar model and the
proto-magnetar model is that the collapsar has to form a black hole.
Being so central to the model, it is legitimate to investigate what
conditions black-hole formation in this context
\citep{woosley:11a}. Surprisingly, little has been done on this
problem.  Numerous simulations so far have focused on the early
collapse phase and the revival of the SN shock, stopping too early to
make any statement concerning black-hole formation.  In 1D, several
studies have investigated the neutrino signal and progenitor
dependence of non-rotating failed SNe \citep{liebendoerfer:04,
  sumiyoshi:07, fischer:09a, oconnor:11}.  \citet{sekiguchi:11a} have
performed 2D simulations of black-hole formation and the subsequent
formation of an accretion disk, although with initial conditions that
are incommensurate with currently suggested LGRB progenitors. In 3D,
\cite{ott:11a} performed fully general-relativistic simulations of
black-hole formation, however using a simplified soft equation of
state that favored it.  Finally, other simulations have started from a
pre-existing \bh\ and investigated the powering of the jet at the
origin of the LGRB \citep{aloy:00,proga:03,zhang:04,lindner:10}, or
the longer-term synthesis of \isoni\ in this unusual context
\citep{milosavljevic:12}.  No simulation has ever demonstrated from
first principles, and thus convincingly, the validity of the collapsar
model, i.e., that the progenitors proposed for this model would indeed
proceed through each and every necessary step: Collapse, formation of
a \pns, failure of the shock revival, formation of a \bh\ followed by
that of a Keplerian disk, and finally the powering of both the LGRB
and the SN, including the synthesis of a generally large amount of
\isoni\ by core-collapse SN standards.  This is an obvious shortcoming
of all theoretical investigations on the collapsar model and its
proposed progenitors.  Similar gaps in the modeling of proto-magnetar
driven LGRBs exist: The early magneto-rotational core-collapse SN
evolution has been modeled in 2D \citep{burrows:07b,takiwaki:11} and
so has the phase in which relativistic outflows are driven
\citep{bucciantini:07,bucciantini:08,bucciantini:09,komissarov:07},
but the evolution connecting the two phases has not been modeled.  The
robustness of the magneto-rotational explosion mechanism largely rests
on the efficiency of angular-momentum transport, and in particular the
extraction of the free-energy stored (and replenished through
accretion) in differential rotation at the surface of the
proto-neutron star. The failure to extract this energy on short time
scales may, however, facilitate black-hole formation, although it may
also compromise energy extraction in the collapsar model.  These
complicated issues require detailed modeling to build upon the
promising results of \citet{thompson:05} and \citet{obergaulinger:09}.
To this day, the collapsar model has been studied more extensively
than the proto-magnetar model for LGRBs, so the later may look more
promising in some ways in part because of the lesser scrutiny it has
received.

In this paper, we focus on one important aspect of the collapsar model
to validate, or invalidate, the assumption, often made but so far
never checked, that the LGRB progenitor models available in the
literature indeed collapse to form a \bh.  We do this by performing
hydrodynamical simulations of the LGRB progenitor models of
\citet{woosley:06} using the code \code{GR1D} \citep{oconnor:10}.
This issue is critical for testing the potential of progenitor stars
for producing LGRBs via the collapsar mechanism, but may also serve to
diagnose an attractive channel for the formation of
proto-magnetars. Such ``failed'' collapsars (because they explode
before forming a \bh) represent a serious alternative for the
production of LGRBs, although they have their own caveats
\citep{metzger:11}.  Admittedly, the phenomenon of core collapse,
bounce, and the events that follow are fundamentally multi-D. We
believe, however, that much can be learned from 1D simulations of the
kind presented here. For example, the mass-accretion rate onto the
\pns\ is largely determined by the angular-averaged density profile of
the progenitor star, which we capture accurately. Our 1D exploration
reveals the landscape of core properties at bounce and quantifies
fundamental differences between progenitors.  In the next section, we
start by reviewing results from stellar-evolution models for LGRB
progenitors.  We then describe our methodology for the \code{GR1D}
simulations of LGRB progenitors available in the literature. We make a
short digression in Section~\ref{sect:rot} to discuss the rotational
properties of the collapsed cores of massive stars.  In
Section~\ref{sect_results}, we present our results before concluding
in Section~\ref{sect_conc}.

\section{Stellar-evolution models of LGRB progenitors}
\label{sect_presn}

Stellar-evolution calculations have been performed to investigate the
mass, rotation, and metallicity requirements for producing
fast-rotating pre-collapse stars. Both single-
\citep{hirschi:05,yoon:05,woosley:06,georgy:09} and binary-star
\citep{petrovic:05,cantiello:07} evolutionary scenarios have been
investigated.  Fast rotation of the proto-stellar core is clearly
critical to procure a large angular momentum to the star initially. If
the rotation rate attained is sufficiently large, the star may even
evolve chemically homogeneously and avoid a supergiant phase, which is
known to sap the core of its rotation through the effects of magnetic
torques. Such fast-rotating chemically-homogeneous stars also
naturally die as H-deficient He-poor Wolf-Rayet (WR) stars.  Low
metallicity quenches the stellar-wind mass loss rate, a condition that
may be more important for a single-star scenario than for the
binary-star scenario \citep{yoon:06,cantiello:07}.  While the general
outcome of these simulations is that it is possible to produce massive
stars with a rapidly spinning core/envelope at death, it is difficult
to compare the final properties of published models. Indeed, models
are rarely evolved all the way to an iron core.  The treatments of
mixing, mass loss, and angular-momentum loss/transport
differ. Magnetic-fields may or may not be included and when they are
the prescription may differ \citep{spruit:02,zahn:07}.

Furthermore, all these studies remain speculative about the outcome of
collapse for such progenitors.  They argue for black hole formation
and the formation of a disk based on order of magnitude estimates,
rather than detailed numerical simulations.  For a start, of all the
LGRB progenitor simulations, only those of \citet{woosley:06} are
evolved all the way to the onset of collapse.  In simulations halted
well before, the iron-core mass is estimated from the CO-core mass
\citep{hirschi:05} or is simply not considered in the discussion
\citep{yoon:05}.  Most studies consider a model viable for producing a
collapsar based exclusively on the angular-momentum budget of the
inner 3\,\msun, and perform no checks on the likelihood of forming a
\bh: its formation is deemed so obvious that the discussion of any
alternate scenario is generally omitted.

Differing in their criteria and approaches for selecting collapsar
candidates, some studies may yield progenitor-mass ranges that do not
even overlap: Using an angular-momentum criterion,
\citet{yoon:05,woosley:06,yoon:06} favor progenitor stars with a
main-sequence mass below $\sim$30\,\msun.  In contrast,
\citet{hirschi:05}, arguing for the need for both a large angular
momentum, a large iron core, and a WO stellar type at death, favor
progenitors with a main-sequence masses above $\sim$35\,\msun\
(magnetic fields are not treated in this work, though).

Recent studies suggest that selecting collapsar progenitors based
exclusively on a large angular-momentum budget may be too simplistic.
For example, the magneto-rotational explosion invoked to explain
hypernovae derives its energy from this same large core angular
momentum (via the MRI and the strongly differentially-rotating layers
in the post-shock region).  This mechanism does not obviously
accommodate the formation of a \bh, as demonstrated by
\citet{dessart:08a}. They simulated the collapse of the core and the
development of a magneto-rotational explosion in model 35OC of
\citet{woosley:06} and found that rotational energy of order 10\,B is
readily available to launch a SN ejecta on a timescale of a few
100\,ms. They furthermore found that accretion is easily shut off by
the developing explosion and that the \pns\ mass fails to grow to the
instability threshold for black-hole formation.  However, the
simulations of \citet{dessart:08a} did not resolve the MRI but instead
used an equipartition ansatz to estimate the magnitude of the
MRI-amplified magnetic fields. In reality, magnetic-field reconnection
may, for example, compromise the dynamical potential of magnetic
stresses, channeling magnetic energy into heat to be radiated away by
neutrinos.  This and other alternatives have been studied by
\citet{thompson:05} under the general form of viscous
dissipation. They find that the extra energy deposition can, in some
cases, considerably alter the post-bounce dynamics and generate a
vigorous explosion.

\citet{oconnor:11} studied black-hole formation based on a variety of
progenitor models characterized by different main-sequence mass,
metallicity, and rotation rate. They find that the outcome of collapse
can be anticipated from the bounce compactness of the progenitor core
structure and in particular that of the region inside 2.5\,\msun,
which corresponds approximately to the maximum mass that a \pns\ can
have and remain in hydrostatic equilibrium.  They find that higher
mass progenitors published in the literature do not always have larger
iron cores and therefore that they are not necessarily more prone to
black-hole formation. They also reveal considerable diversity in
progenitor core structure, even for the same main-sequence mass. Some
stellar evolution studies obtain a monotonic increase of the iron-core
mass (or bounce compactness; see Fig.~9 of \citealt{oconnor:11}) versus
main-sequence mass (\citealt{limongi:06}; see also
\citealt{hirschi:04,hirschi:05}), while some show an anti-correlation
beyond $\sim$40\,\msun\ \citep{woosley:07}. The primary
  reason for this are differing prescriptions for rate and time of
  mass loss, one of the major uncertainties in massive star evolution
  (see also the discussions in \citealt{hirschi:05} and
  \citealt{oconnor:11}).

\begin{deluxetable*}{l@{\hspace{-11mm}}cc|ccc|cccc|ccc|c}
\tabletypesize{\scriptsize}
\tablecolumns{14}
\tablewidth{0pc}
\tablecaption{Progenitor Model Properties}
\tablehead{Model\tablenotemark{a} & $\xi_{2.5}$ &
  $\Omega_c$\tablenotemark{b} & $t_\mathrm{BH}$\tablenotemark{c} & $M_\mathrm{b, max}$\tablenotemark{d}
  & $M_\mathrm{g, max}$\tablenotemark{e}  &
  $M_\mathrm{b,BH}^\mathrm{DF}$\tablenotemark{f} &
  $a^\star_\mathrm{BH,DF}$\tablenotemark{g} &
  $t_\mathrm{DF}$\tablenotemark{h} & $M_\mathrm{preSN} -
  M_\mathrm{b,BH}^\mathrm{DF}$\tablenotemark{i} &
  $\bar{\eta}_\mathrm{heat}^\mathrm{crit}$ & $P_\text{ref}$\tablenotemark{j} &
  $I$\tablenotemark{k} &$F_\mathrm{rot}^\mathrm{100ms}$~~\tablenotemark{l} \\
&&(s$^{-1}$) & (s)&($M_\odot$) &($M_\odot$) & ($M_\odot$) & & (s) &
($M_\odot$) & & (ms) & ($10^{45}$\,g\,cm$^2$) & (B)}
\startdata
12SA & 0.003 & 0.000 & $\cdots$  & (1.56) & (1.47) & (10.9)\phm{0} & $\cdots$ & $\cdots$           & $\cdots$           & 0.167 & $\cdots$    & $\cdots$ & $\cdots$ \\
12SG & 0.239 & 0.198 & 2.728     & 2.33   & 2.12   & \phm{0}(7.57) & $\cdots$ & $\cdots$           & $\cdots$           & 0.124 & 18.1\phm{0} & 3.37     & 0.034    \\
12SH & 0.141 & 0.144 & $\cdots$  & (2.08) & (1.91) & \phm{0}(5.43) & $\cdots$ & $\cdots$           & $\cdots$           & 0.148 & 25.5\phm{0} & 3.35     & 0.008    \\
12SI & 0.075 & 0.208 & $\cdots$  & (1.75) & (1.64) & \phm{0}(6.95) & $\cdots$ & $\cdots$           & $\cdots$           & 0.153 & 24.7\phm{0} & 3.17     & 0.010    \\
12SJ & 0.121 & 0.751 & $\cdots$  & (2.05) & (1.90) & \phm{0}6.77   & 0.470    & \phm{0}91.6\phm{0} & \phm{0}2.27        & 0.166 & \phm{0}4.10 & 3.53     & 0.305    \\
\hline
12OA & 0.011 & 0.000 & $\cdots$  & (1.52) & (1.44) & (11.9)\phm{0} & $\cdots$ & $\cdots$           & $\cdots$           & 0.104 & $\cdots$    & $\cdots$ & $\cdots$ \\
12OG & 0.029 & 0.149 & $\cdots$  & (1.88) & (1.75) &  \phm{0}4.50  & 0.217    & 4.42$\times10^6$   & \phm{0}7.32        & 0.179 & 22.5\phm{0} & 3.17     & 0.006    \\
12OH & 0.090 & 0.285 & $\cdots$  & (1.84) & (1.71) &  \phm{0}7.62  & 0.210    & 710.\phm{00}       & \phm{0}0.07        & 0.135 & 17.1\phm{0} & 3.30     & 0.020    \\
12OI & 0.095 & 1.061 & $\cdots$  & (1.86) & (1.73) &  \phm{0}5.91  & 0.535    & \phm{0}68.8\phm{0} & \phm{0}3.81        & 0.181 & \phm{0}4.49 & 3.27     & 0.270    \\
12OL & 0.076 & 0.299 & $\cdots$  & (1.75) & (1.64) &  \phm{0}6.99  & 0.259    & 554.\phm{00}       & \phm{0}0.36        & 0.136 & 19.7\phm{0} & 3.61     & 0.017    \\
12ON & 0.170 & 1.709 & $\cdots$  & (2.21) & (2.02) &  \phm{0}2.67  & 0.496    & \phm{00}8.51       & \phm{0}8.26        & 0.145 & \phm{0}2.38 & 3.27     & 1.013    \\
\hline
12TA & 0.008 & 0.000 & $\cdots$  & (1.59) & (1.49) & (12.0)\phm{0} & $\cdots$ & $\cdots$           & $\cdots$           & 0.117 & $\cdots$    & $\cdots$ & $\cdots$ \\
12TG & 0.034 & 0.148 & $\cdots$  & (1.91) & (1.77) &  \phm{0}4.59  & 0.228    & 3.73$\times10^6$   & \phm{0}7.35        & 0.182 & 24.9\phm{0} & 3.43     & 0.007    \\
12TH & 0.107 & 1.042 & $\cdots$  & (1.93) & (1.79) &  \phm{0}6.67  & 0.495    & \phm{0}85.2\phm{0} & \phm{0}2.56        & 0.138 & \phm{0}5.06 & 3.59     & 0.313    \\
12TI & 0.145 & 1.323 & $\cdots$  & (2.02) & (1.86) &  \phm{0}3.33  & 0.507    & \phm{0}15.3\phm{0} & \phm{0}7.46        & 0.144 & \phm{0}3.26 & 3.48     & 0.610    \\
12TJ & 0.517 & 1.281 & 0.853     & 2.51   & 2.37   &  \phm{0}2.97  & 0.640    & \phm{00}2.64       & \phm{0}8.58        & 0.191 & \phm{0}1.10 & 4.31     & 3.286    \\
\hline
16SA & 0.101 & 0.000 & $\cdots$  & (1.88) & (1.74) & (14.6)\phm{0} & $\cdots$ & $\cdots$           & $\cdots$           & 0.138 & $\cdots$    & $\cdots$ & $\cdots$ \\
16SG & 0.109 & 0.203 & $\cdots$  & (1.91) & (1.77) &  \phm{0}9.79  & 0.404    & 2.71$\times10^7$   & \phm{0}2.16        & 0.141 & 20.8\phm{0} & 3.27     & 0.010    \\
16SH & 0.081 & 0.341 & $\cdots$  & (1.76) & (1.64) & \phm{0}(7.70) & $\cdots$ & $\cdots$           & $\cdots$           & 0.182 & 16.8\phm{0} & 2.68     & 0.019    \\
16SI & 0.380 & 0.189 & 1.132     & 2.38   & 2.20   & \phm{0}(9.85) & $\cdots$ & $\cdots$           & $\cdots$           & 0.158 & 11.9\phm{0} & 3.82     & 0.062    \\
16SL & 0.075 & 0.207 & $\cdots$  & (1.73) & (1.62) & \phm{0}(6.30) & $\cdots$ & $\cdots$           & $\cdots$           & 0.130 & 28.4\phm{0} & 3.31     & 0.007    \\
16SM & 0.121 & 0.229 & $\cdots$  & (2.02) & (1.85) & \phm{0}(8.31) & $\cdots$ & $\cdots$           & $\cdots$           & 0.145 & 20.0\phm{0} & 3.51     & 0.016    \\
16SN & 0.496 & 0.455 & 0.777     & 2.42   & 2.27   & \phm{0}9.45   & 0.508    & \phm{0}40.6\phm{0} & \phm{0}1.77        & 0.187 & \phm{0}3.25 & 3.69     & 0.451    \\
\hline
16OA & 0.144 & 0.000 & $\cdots$  & (2.15) & (1.96) & (15.8)\phm{0} & $\cdots$ & $\cdots$           & $\cdots$           & 0.133 & $\cdots$    & $\cdots$ & $\cdots$ \\
16OG & 0.193 & 0.176 & 3.437     & 2.33   & 2.11   &  \phm{0}7.16  & 0.230    & 4.37$\times10^6$   & \phm{0}8.49        & 0.168 & 17.6\phm{0} & 3.82     & 0.018    \\
16OH & 0.185 & 0.248 & $\cdots$  & (2.21) & (2.00) &  \phm{0}9.18  & 0.133    & 810.\phm{00}       & 7.9$\times10^{-5}$ & 0.150 & 20.5\phm{0} & 3.59     & 0.023    \\
16OI & 0.344 & 0.733 & 1.449     & 2.37   & 2.19   &  \phm{0}7.10  & 0.553    & \phm{0}18.2\phm{0} & \phm{0}5.11        & 0.152 & \phm{0}3.21 & 4.06     & 0.700    \\
16OL & 0.124 & 0.316 & $\cdots$  & (2.02) & (1.86) &  \phm{0}8.63  & 0.200    & 593.\phm{00}       & \phm{0}0.05        & 0.138 & 14.1\phm{0} & 3.41     & 0.030    \\
16OM & 0.172 & 1.059 & $\cdots$  & (2.17) & (1.98) &  \phm{0}5.64  & 0.590    & \phm{0}24.5\phm{0} & \phm{0}6.31        & 0.177 & \phm{0}2.60 & 3.02     & 0.480    \\
16ON & 0.357 & 1.382 & 1.458     & 2.40   & 2.22   &  \phm{0}3.36  & 0.582    & \phm{00}4.70       & 10.8\phm{0}        & 0.162 & \phm{0}1.40 & 3.59     & 2.082    \\
\hline
16TA & 0.070 & 0.000 & $\cdots$  & (1.76) & (1.64) & (16.0)\phm{0} & $\cdots$ &  $\cdots$          & $\cdots$           & 0.148 & $\cdots$    & $\cdots$ & $\cdots$ \\
16TG & 0.288 & 0.242 & 1.738     & 2.35   & 2.16   & 13.3\phm{0}   & 0.366    & 3.44$\times10^4$   & \phm{0}2.41        & 0.174 & 10.4\phm{0} & 4.02     & 0.083    \\
16TH & 0.434 & 0.598 & 0.958     & 2.41   & 2.25   &  \phm{0}8.01  & 0.511    & \phm{0}23.7\phm{0} & \phm{0}3.44        & 0.151 & \phm{0}2.57 & 3.80     & 0.599    \\
16TI & 0.242 & 1.367 & 2.791     & 2.41   & 2.21   &  \phm{0}3.51  & 0.554    & \phm{0}10.6\phm{0} & 10.4\phm{0}        & 0.150 & \phm{0}2.17 & 3.77     & 1.341    \\
\hline
35OA & 0.178 & 0.289 & $\cdots$  & (2.26) & (2.05) & (12.9)\phm{0} & $\cdots$ & $\cdots$           & $\cdots$           & 0.153 & 14.6\phm{0} & 3.53     & 0.044    \\
35OB & 0.537 & 1.545 & 0.776     & 2.42   & 2.25   & 16.4\phm{0}   & 0.545    & \phm{0}31.5\phm{0} & \phm{0}4.80        & 0.198 & \phm{0}1.44 & 3.48     & 3.617    \\
35OC & 0.458 & 1.980 & 0.972     & 2.43   & 2.29   &  \phm{0}4.44  & 0.622    & \phm{00}4.84       & 23.6\phm{0}        & 0.162 & \phm{0}1.10 & 4.49     & 7.521    \\
\hline
HE16C & 0.137 & 0.133 & $\cdots$ & (2.06) & (1.90) & \phm{0}(5.15) & $\cdots$ & $\cdots$           & $\cdots$           & 0.134 & 25.5\phm{0} & 3.11     & 0.007    \\
HE16D & 0.283 & 0.440 & 1.706    & 2.35   & 2.16   &  \phm{0}8.65  & 0.367    & 117.\phm{00}       & \phm{0}0.88        & 0.171 & \phm{0}4.74 & 3.65     & 0.206    \\
HE16E & 0.129 & 1.428 & $\cdots$ & (2.00) & (1.85) &  \phm{0}6.82  & 0.594    & \phm{0}43.1\phm{0} & \phm{0}6.05        & 0.153 & \phm{0}3.77 & 3.27     & 0.447    \\
HE16F & 0.496 & 1.096 & 0.837    & 2.48   & 2.33   &  \phm{0}4.29  & 0.567    & \phm{00}8.97       & 10.5\phm{0}        & 0.202 & \phm{0}1.46 & 4.28     & 2.323    \\
HE16H & 0.610 & 1.196 & 0.641    & 2.52   & 2.38   &  \phm{0}4.12  & 0.597    & \phm{00}5.95       & 11.6\phm{0}        & 0.204 & \phm{0}1.26 & 4.23     & 3.335    \\
HE16K & 0.132 & 0.134 & $\cdots$ & (2.04) & (1.88) & \phm{0}(5.16) & $\cdots$ & $\cdots$           & $\cdots$           & 0.140 & 26.6\phm{0} & 3.11     & 0.007    \\
HE16L & 0.316 & 0.315 & 1.497    & 2.36   & 2.17   &  \phm{0}9.34  & 0.286    & 195.\phm{00}       & \phm{0}0.24        & 0.165 & \phm{0}6.73 & 3.73     & 0.116    \\
HE16M & 0.111 & 1.206 & $\cdots$ & (1.91) & (1.77) & 10.4\phm{0}   & 0.532    & \phm{0}94.0\phm{0} & \phm{0}2.61        & 0.134 & \phm{0}4.42 & 3.23     & 0.348    \\
HE16N & 0.198 & 1.203 & 3.424    & 2.36   & 2.15   &  \phm{0}7.81  & 0.604    & \phm{0}35.9\phm{0} & \phm{0}7.15        & 0.154 & \phm{0}2.59 & 3.84     & 0.970    \\
HE16O & 0.298 & 1.209 & 1.891    & 2.39   & 2.20   &  \phm{0}6.77  & 0.594    & \phm{0}25.0\phm{0} & \phm{0}8.86        & 0.127 & \phm{0}1.98 & 3.61     & 1.414    \\
HE16P & 0.573 & 1.038 & 0.672    & 2.49   & 2.35   &  \phm{0}6.42  & 0.584    & \phm{0}14.2\phm{0} & \phm{0}9.46        & 0.235 & \phm{0}1.58 & 4.16     & 2.523
\enddata
\tablenotetext{a}{Model designation from \citet{woosley:06}. See text
for details.}
\tablenotetext{b}{Initial central angular velocity.}
\tablenotetext{c}{Time elapsed between bounce and black hole formation, $\cdots$ indicates that no black hole formed within 3.5\,s of bounce.}
\tablenotetext{d}{Baryonic mass of the \pns\ at the time of black hole formation.  If no black hole forms in 3.5\,s, we give the \pns\ baryonic mass at 3.5\,s.}
\tablenotetext{e}{Gravitational mass of the \pns\ at the time of black hole formation.  If no black hole forms in 3.5\,s, we give the \pns\ gravitational mass at 3.5\,s.}
\tablenotetext{f}{Baryonic mass interior to the \isco at the time of disk formation. If the angular momentum is too low to foster disk formation, we give the progenitor mass in parentheses instead.}
\tablenotetext{g}{Dimensionless spin of the black hole when the disk forms. $\cdots$ indicates no disk forms.}
\tablenotetext{h}{Twice the free-fall time of the mass element at the \isco. $\cdots$ indicates no disk forms.}
\tablenotetext{i}{Baryonic mass outside of the black hole at disk formation.  $\cdots$ indicates no disk forms.}
\tablenotetext{j}{Rotational period $P_\text{ref}=2\pi/\bar{\omega}$ computed by appoximating $\bar{\omega}$ as
the ratio of the total angular momentum to the moment of inertia of the \pns, Eq.~\ref{eq:omegabar}. $\cdots$ indicates a non-rotating model.}
\tablenotetext{k}{Proto-neutron star moment of inertia, note this can be up to two times the value for a non-rotating cold neutron star.
This is the origin of the discrepancy with the values presented by \citet{woosley:06},
who consider a cold non-rotating 1.4\,\msun\ neutron star with a moment of inertia $I = 1.4\times10^{45}\,\mathrm{g}\,\mathrm{cm}^2$. $\cdots$ indicates a non-rotating model.}
\tablenotetext{l}{Free energy stored in differential rotation. This amounts
 to the energy difference between the \pns\ we obtain with \code{GR1D} and the corresponding \pns\ with the same total angular
 momentum and moment of inertia but assuming solid-body rotation. $\cdots$ indicates a non-rotating model.}
\label{tab:results}
\end{deluxetable*}

\section{Methods \& Initial Model Set}
\label{sect_method}

In this work, we use the open-source, spherically symmetric, general
relativistic, Eulerian hydrodynamics code \code{GR1D}
\citep{oconnor:10}.  Rotation is included through a
centrifugal-acceleration term in the momentum equation --- this is the
most important dynamical feature of rotation relevant to core
collapse. However, \code{GR1D} cannot account for the associated
deviations from spherical symmetry nor any angular-momentum
redistribution.  We select the equation of state (EOS) from
\citet{lseos:91} characterized by a nuclear incompressibility of
$220\,$MeV (hereafter referred to as the LS220 EOS).  This EOS
provides the best match to both mass and mass-radius constraints from
observations and nuclear theory
\citep{demorest:10,oezel:10b,steiner:10,hebeler:10}. \code{GR1D} uses
an efficient neutrino leakage/heating scheme that qualitatively
reproduces the salient features of neutrino transport. We refer the
reader to \citet{oconnor:10,oconnor:11} for additional details on
\code{GR1D} and our methodology.

As described above, the only stellar-evolutionary models for LGRB
progenitors that are evolved until the onset of collapse are those
proposed by \citet{woosley:06}.  We thus focus on their model dataset
for our investigation on the dynamics of the core-collapse SN
engine and the potential formation of a black hole in the collapsar
context.  Using \code{KEPLER}, \citet{woosley:06} investigated a
rather narrow range of progenitor masses, but varied the initial
rotation rate (solid-body rotation is assumed initially) and
environmental metallicity from solar to 1\% solar (with an additional
tunable factor as low as 0.1 for the metallicity-dependent mass loss
rate, equivalent to a reduction in metallicity by a factor of
100 in their mass-loss prescription).  Arguing that the inferred
mass of LGRB/SN ejecta known in 2006 is on the order of 10\,\msun, and since
higher-mass stars may lose too much angular momentum through stellar
winds (even at low metallicity), they focused primarily on lower-mass
progenitors, with main-sequence masses of 12 and 16\,\msun,\footnote{They
  also perform simulations for 16-\msun\ helium cores and find
  comparable outcomes.} with the exception of one 35\,\msun\ model
set.

We adopt the same nomenclature as for their 12-, 16- and 35-$M_\odot$
models. It comprises the model's main-sequence mass, followed by a
letter denoting the environmental metallicity (`S' for solar, `O' for
10\% solar, and `T' for 1\% solar).  An additional letter is appended
to individualize the models done with different WR mass-loss rate
prescriptions, allowing or not for magnetic effects, and the total
angular momentum of the star. 16-$M_\odot$ helium models are denoted
by `HE16' followed by an individualizing capital letter.

In this work, we simulate the collapse and post-bounce evolution with
\code{GR1D} for all these progenitor models, with a primary focus on
determining their ability to produce the key features of the collapsar
model: A \bh\ together with a Keplerian disk.  As we
  discuss in the following section, black-hole formation is not obviously
  guaranteed in any of these dying stars.

\section{Notes on Rotating Core Collapse}
\label{sect:rot}

Since LGRBs seem fundamentally related to rapid rotation, it is useful
to summarize a few facts and concepts related to the gravitational
collapse of rotating iron cores in massive stars.  First, it is
reasonable to assume (which is borne out by simulations, e.g.,
\citealt{heger:05}) that the iron core, in its pre-collapse state,
will be approximately uniformly rotating. Such a solid-body rotation
corresponds to the lowest energy state at fixed total angular momentum
and will be assumed on a secular timescale by any rotating fluid that
has some means to redistribute angular momentum.

\begin{figure}[t!]
\includegraphics[width=0.97\columnwidth]{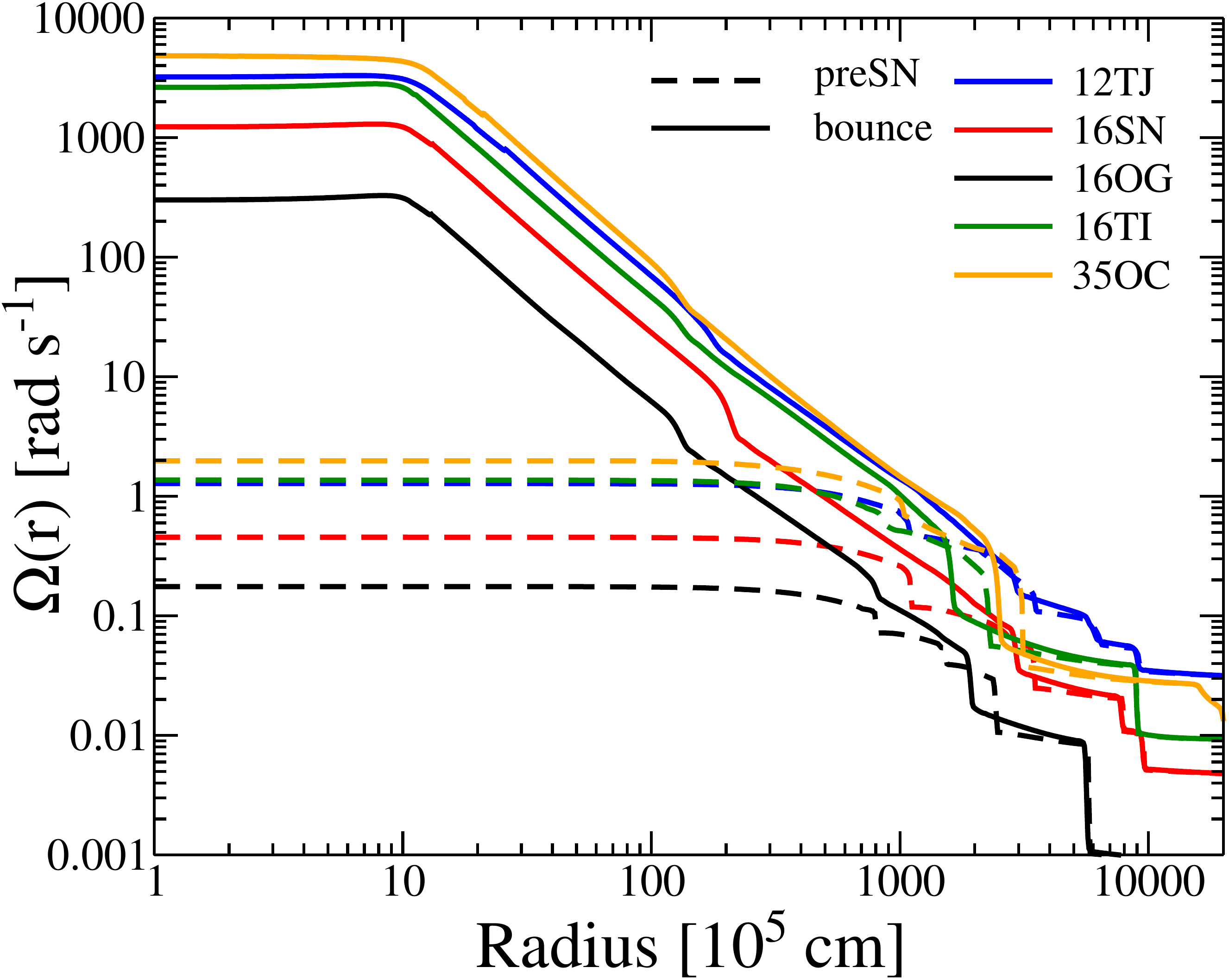}
\caption{Angular velocity $\Omega(r)$ versus radius $r$ at both the
  pre-SN stage (dashed lines) and at core bounce (solid lines)
  for selected models of \cite{woosley:06}.  The inner homologously
  collapsing core maintains its initial uniform rotation throughout
  collapse. \label{fig:omega} }
\end{figure}

Rotating core collapse, even for the high pre-collapse rotation rates
of some of the potential LGRB progenitors that we consider in this
study, proceeds qualitatively in a very similar fashion to
non-rotating collapse as long as the ratio of the centrifugal
acceleration $a_\text{cent}$ to the gravitational acceleration
$a_\text{grav}$, is small,
\begin{equation}
\frac{a_\mathrm{cent}}{a_\mathrm{grav}} = \frac{\Omega^2(r) r}{G M(r) r^{-2}} = \frac{\Omega^2(r) r^3}{G M(r)} \ll 1\,\,.
\end{equation}
Due to angular-momentum conservation, the angular velocity behaves as
$\Omega(r) \propto r^{-2}$. $M(r)$ stays constant for a collapsing
mass shell, and, thus, the above ratio increases during collapse as
$r^{-1}$ and may potentially become large for small radii.

In the case of $a_\mathrm{cent}/ a_\mathrm{grav} \ll 1$, the
collapsing rotating iron core will behave like a non-rotating core and
separate into a subsonically collapsing inner core ($|v_r(r)| <
c_s(r)$) and a supersonically collapsing outer core ($|v_r(r)| >
c_s(r)$). The inner core exhibits a self-similar (homologous) velocity
profile, $v(r) \propto r$, until core bounce and shock formation
\citep{goldreich:80}. After core bounce, the inner core material forms
the core of the \pns\ and outer core material accumulates at its edge.
The mass of the inner core at bounce is typically $\sim$$0.5\,M_\odot$
for non-rotating cores, set by the EOS of relativistic electrons, the
trapped lepton number, core entropy, and gravity \citep{burrows:83}.
It increases monotonically, though slowly, with increasing pre-collapse
rotation rate. For cores that reach $a_\mathrm{cent} / a_\mathrm{grav}
\approx 1$, the mass of the inner core will be increased to $\gtrsim
0.7\,M_\odot$ \citep{dimmelmeier:08}.

Since the inner core is collapsing homologously, we can introduce
a homology parameter $\alpha(t)$, so that
\begin{equation}
r(t) = \alpha(t)\, r_0\,\,,\label{eq:selfsimilar}
\end{equation}
where $r(t)$ is the radius of a collapsing fluid element at time $t$
and $r_0$ is its initial radius. This must hold for any mass shell
within the inner core.  Two collapsing fluid elements
  located initially at $r_1$ and $r_2$, and rotating with a frequency
  $\Omega_1 = j_1/r_1^2$ and $\Omega_2 = j_2/r_2^2$ conserve angular
  momentum, $\Omega'_1 {r'_1}^2 = j'_1 = j_1 = \Omega_1r_1^2$.
  Homology implies $r'_1 = \alpha(t) r_1$, and $r'_2 = \alpha(t) r_2$, therefore
\begin{equation}
\frac{\Omega_1'(r_1')}{\Omega'_2(r'_2)} = \frac{j_1 / {r'_1}^2}{j_2/{r'_2}^2} =
\frac{j_1}{j_2}\frac{\alpha(t)^2 r_2^2}{\alpha(t)^2 r_1^2} = \frac{\Omega_1}{\Omega_2} \,\,.
\end{equation}
Since this property holds for any mass shell within the inner core,
the rotational profile must be preserved under homologous collapse. In
practice, because the pre-collapse cores of massive stars are always
in solid-body rotation, so are the inner cores of the \pns\ at bounce.
Outside of the homologously collapsing core the self similar relation
of Eq.~\ref{eq:selfsimilar} does not hold. Most generally, gradients
in the rotation rate develop due to the underlying density gradients
in the hot postshock region and in the supersonically infalling region
ahead of the accretion shock.

Based on these arguments, we expect an early post-bounce rotational
profile that is approximately uniform within the inner
$0.5-0.7\,M_\odot$ (out to $\sim$$10-15\,\mathrm{km}$ in radius) and
strongly differential at larger radii. This is confirmed by
Fig.~\ref{fig:omega} which shows $\Omega(r)$ at bounce as obtained
with \code{GR1D} for a variety of models considered in this study. We
also show the rotational profile at the onset of collapse. This result
is not entirely new, but has previously been pointed out by
\cite{ott:06spin} in the context of 1D and 2D rotating core collapse
simulations.

Since uniform rotation is the lowest energy state, the \emph{shear
  energy} of differential rotation is to be interpreted as a free
energy that will be tapped by any process (e. g., nonaxisymmetric
rotational shear instabilities, viscosity, or the MRI) capable of
redistributing angular momentum. Viscosity would lead to additional
heating in the postshock region to enhance the neutrino mechanism
\citep{thompson:05} while the MRI action could strengthen the magnetic
fields, driving bipolar outflows in the magneto-rotational mechanism
\citep{burrows:07b}.

In our simulations, we estimate the available free energy of
differential rotation by computing the difference in rotational energy
of the \pns\ model in \code{GR1D} and the rotational energy of a
uniformly spinning \pns\ of the same angular momentum and moment of
inertia,
\begin{equation}
  F_\text{rot} = T - \frac{I\bar{\omega}^2}{2}\,,
\label{eq:freeenergy}
\end{equation}
\noindent
where from \cite{oconnor:10}, for \code{GR1D},
\begin{eqnarray}
T &=&  {4 \pi \over 3} \int_0^{R_\text{PNS}} \rho h X W^2 v^2_\varphi r^2 dr\,\,,\\
I &=&  {8 \pi \over 3} \int_0^{R_\text{PNS}} \rho h X W^2 r^4 dr\,\,,\\
\bar{\omega} &= & \int_0^{R_\text{PNS}} \rho h X W^2 rv_\varphi r^2 dr \bigg/
\int_0^{R_\text{PNS}} \rho h X W^2 r^4 dr\label{eq:omegabar}
\end{eqnarray}
\noindent
and $T$ is the rotational energy, $I$ is the moment of inertia, and
$\bar{\omega}$ is the uniform rotation frequency, $h$ is the specific
enthalpy, $X^2$ is the $g_{rr}$ component of the metric, $W$ is the
Lorentz factor, and $v_\varphi$ is the angular velocity. We take
$R_\text{PNS}$ to be the radius where the matter density, $\rho =
10^{10}\,$g\,cm$^{-3}$.

\section{Results}
\label{sect_results}

We have performed simulations for the entire set of \code{KEPLER}
models published in \citet{woosley:06}\footnote{\url{Models are
    available from http://homepages.spa.umn.edu/\textasciitilde
    alex/GRB2/}}.  We first consider the rapidly spinning progenitors
evolved without magnetic fields.  All these models have a
dimensionless Kerr spin ($a^\star = Jc/GM^2$) at 3\,\msun\ greater
than unity (with the exception of model HE16J, which has $a^\star =
0.91$) and are thus considered as promising collapsar candidates by
\citet{woosley:06}.  Unfortunately, when evolved with \code{GR1D}, the
collapsing iron core of all such models halts its collapse and expands
--- these models do not experience core bounce within a few seconds of
evolution in \code{GR1D}.  We associate this problem with the neglect
of the centrifugal acceleration in the momentum equation in
\code{KEPLER}, an approximation that fails in the fastest rotating
models. This term is included in \code{GR1D}. The mismatch suggests
that their fastest models may be significantly affected by the
addition of this term. Even if they did collapse, it is not clear that
such extremely fast rotating cores would avoid a centrifugal bounce.

In the remainder of this paper, we thus limit our discussion to models
evolved with magnetic fields and therefore subject to magnetic torques
during their evolution.  Of the 46 models that fulfill this criterion,
we identify 4 additional models (12OM, 16TJ, 35OD, and HE16G) that do
not collapse but instead expand when restarted with \code{GR1D}. We
exclude these as well from our study.  Finally, for reference and
completeness, we include the non-rotating models associated with each
series (12SA, 12OA, 12TA, 16SA, 16OA, 16TA), making a total of 48
models.  Each simulation is continued after core bounce until a
\bh\ forms or until a time of $3.5\,$s has passed, whichever
comes first.  We present the results for these 48 models evolved with
magnetic fields in Table~\ref{tab:results} (for the table layout, we
group models in bundles first of increasing mass, then of decreasing
metallicity, and finally in alphabetical order which generally
corresponds to an increased initial rotation rate).

As advocated by \citet{oconnor:11}, the bounce compactness is a robust
quantity for diagnosing the propensity to black-hole formation, which
is largely determined by the spatial extent encompassed by the
2.5\,\msun\ Lagrangian mass coordinate at core bounce in the
progenitor core.  To avoid introducing biases associated with the
non-uniform conditions in the progenitor simulations (\code{KEPLER}
models are not all evolved to the same central density on their
collapse trajectory) this compactness is unambiguously evaluated at
the time of bounce.  The formal definition of this core compactness is
\begin{equation}
  \xi_{M} = {M / \,M_\odot  \over R(M_\mathrm{bary} =
    M) / 1000\,\mathrm{km}}\Big|_{t =t_{\mathrm{bounce}}}\,,\label{eq:bouncecompactness}
\end{equation}
where we take $M=2.5\,M_\odot$.
$R(M_\mathrm{bary}=2.5\,M_\odot)$ is the radial coordinate that
encloses 2.5\,$M_\odot$ of baryonic material at the time of core bounce
\citep{oconnor:11}.

\begin{figure}[htbp]
\includegraphics[width=0.97\columnwidth]{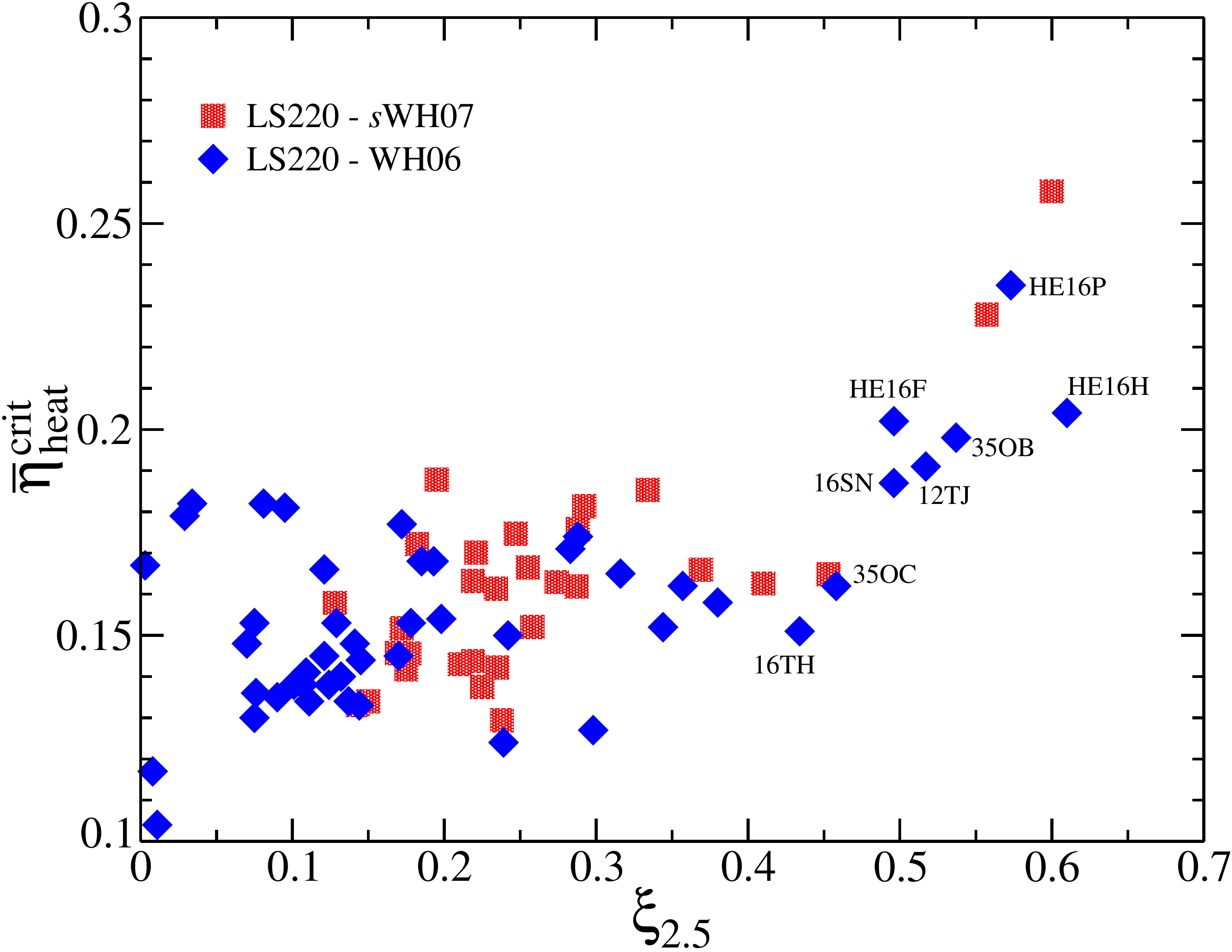}
\caption{ Illustration of the critical heating efficiency
  $\bar{\eta}_\mathrm{heat}^\mathrm{crit}$ versus bounce compactness
  $\xi_{2.5}$ for our \code{GR1D} simulations of the
  \citet{woosley:06} models, whose properties are summarized in
  Table~1 (blue diamonds). For comparison, we overplot the same
  quantity for the standard non-rotating core-collapse SN progenitor
  models of \citet{woosley:07} evolved at solar metallicity (red
  squares).  For the most part, the two distributions overlap,
  suggesting that the propensity to black-hole formation and explosion
  is comparable for both.  Only models with the fastest rotation rates
  achieve a larger compactness in excess of 0.4--0.5, but these may
  then be diverted from black-hole formation through an early
  magneto-rotational explosion.
\label{fig:eta_xi}
}
\end{figure}

Our simulations first demonstrate that most of the models have a small
core compactness $\xi_{2.5}$.  \citet{oconnor:11} argues that a
compactness of 0.45 represents a threshold value, for the neutrino
mechanism, since above it an unrealistic neutrino-heating
efficiency is required to prevent black-hole formation. We further
confirm this by determining the critical heating efficiency for the
models in Table~\ref{tab:results} via the same procedure as in
\citet{oconnor:11}, to which we refer the reader for full details. We
note that this criterion is for explosions via the neutrino mechanism
and therefore neglects any magneto-rotational contribution to the
powering of an explosion \citep{dessart:08a}, so this threshold value
is probably a lower limit.  We also stress that this criterion is
based on spherically symmetric simulations with an efficient, but
crude, approximation to neutrino transport.  Therefore we
  suggest caution when attempting to interpret the outcome of a model
  based solely on the bounce compactness, since the adopted threshold
  value for BH formation is a semi-quantitative estimate based on
  where the slope of the required heating efficiency begins to clearly
  increase (see the discussion in \citealt{oconnor:11}). When
plotting the critical heating efficiency of the non-rotating solar
metallicity stars from \citet{woosley:07} determined in
\citet{oconnor:11}, together with the generally fast-rotating
progenitors of \citet{woosley:06}, we find that both datasets in fact
overlap for the most part (Fig.~\ref{fig:eta_xi}).  In other words, in
terms of compactness, most of these progenitors are similar to garden
variety, low-mass, non-rotating, progenitors and do not seem to have
any more reason to form a \bh\ than, e.g., the RSG progenitors
expected to produce SNe II-Plateau.  As shown in
Table~\ref{tab:results}, provided no explosion is launched, half of
these models have not formed a \bh\ after 3.5\,s, and only $\sim$15\%
do within $\sim$1\,s. We note that this result is not so surprising
given the small iron-core mass ($\sim 1.4\,M_\odot$) of most
\citet{woosley:06} models (see their Tables~1 and 2).  The first
conclusion from our work is therefore that most of the models
presented here are rather unlikely to form a black hole and thus may
fail in a very fundamental way to produce a collapsar, irrespective of
their angular-momentum budget.

Within each sequence presented in Table~\ref{tab:results}, the models
that form a \bh\ within 3.5\,s of core bounce, and thus at least in
principle susceptible to form a collapsar, are the faster rotating
ones characterized by very weak mass-loss rates.  These properties
conspire to produce larger CO cores, more typical of more massive
stars that do not rotate. In the following discussion, we group these
models into several categories.

The first category are models which obviously do not give rise to a
LGRB, either by the standard Type-I collapsar or the proto-magnetar
mechanism, because they contain too little angular momentum.
Optimistically assuming a failed core-collapse SN, which is
unlikely given the modest values of $\xi_{2.5}$, models 12SG
($\xi_{2.5} = 0.239$), 16OG ($\xi_{2.5} = 0.193$), 16SI ($\xi_{2.5} =
0.380$), and 16TG ($\xi_{2.5} = 0.288$)  possess too
little angular momentum in the remainder of the star to form a disk
about the central black hole within $10^6$\,s of collapse.  This
behavior is reflected by the stellar type at the time of death, i.e. a
BSG star for model 12SG and a RSG star for models 16OG and 16TG, only
16SI is a WR star at the time of death. Quantitatively, this can be
further inspected in Table~\ref{tab:results} where we include the disk
formation time, the black hole mass and spin at that time, and the mass
exterior to the disk. We define disk formation
to be when the accreting material will first be supported at the
\isco\ about a black hole with the enclosed mass and angular momentum
using the formulae of \citet{bardeen:72}.  We estimate the disk
formation time as twice the free fall time of the innermost mass element
that reaches a Keplerian velocity \citep{oconnor:11, burrows:86bh}.
\begin{equation}
t_\text{DF} = 2\times \pi \sqrt{\frac{[r_\text{pre-SN}(M_\text{disk})]^3}{8GM_\text{disk}}}
\end{equation}
\noindent
where $r_\text{pre-SN}(M_\text{disk})$ is the radius of the
disk-forming Lagrangian mass element in the pre-SN model. If no such
mass element exists, no disk will form.  In this case we include,
instead of the enclosed black-hole mass, the total pre-SN stellar mass
in parentheses. In the four models mentioned above, either no disk
forms or the disk formation time is $\gtrsim 10^6\,$s.

\begin{figure}[htbp]
\includegraphics[width=0.97\columnwidth]{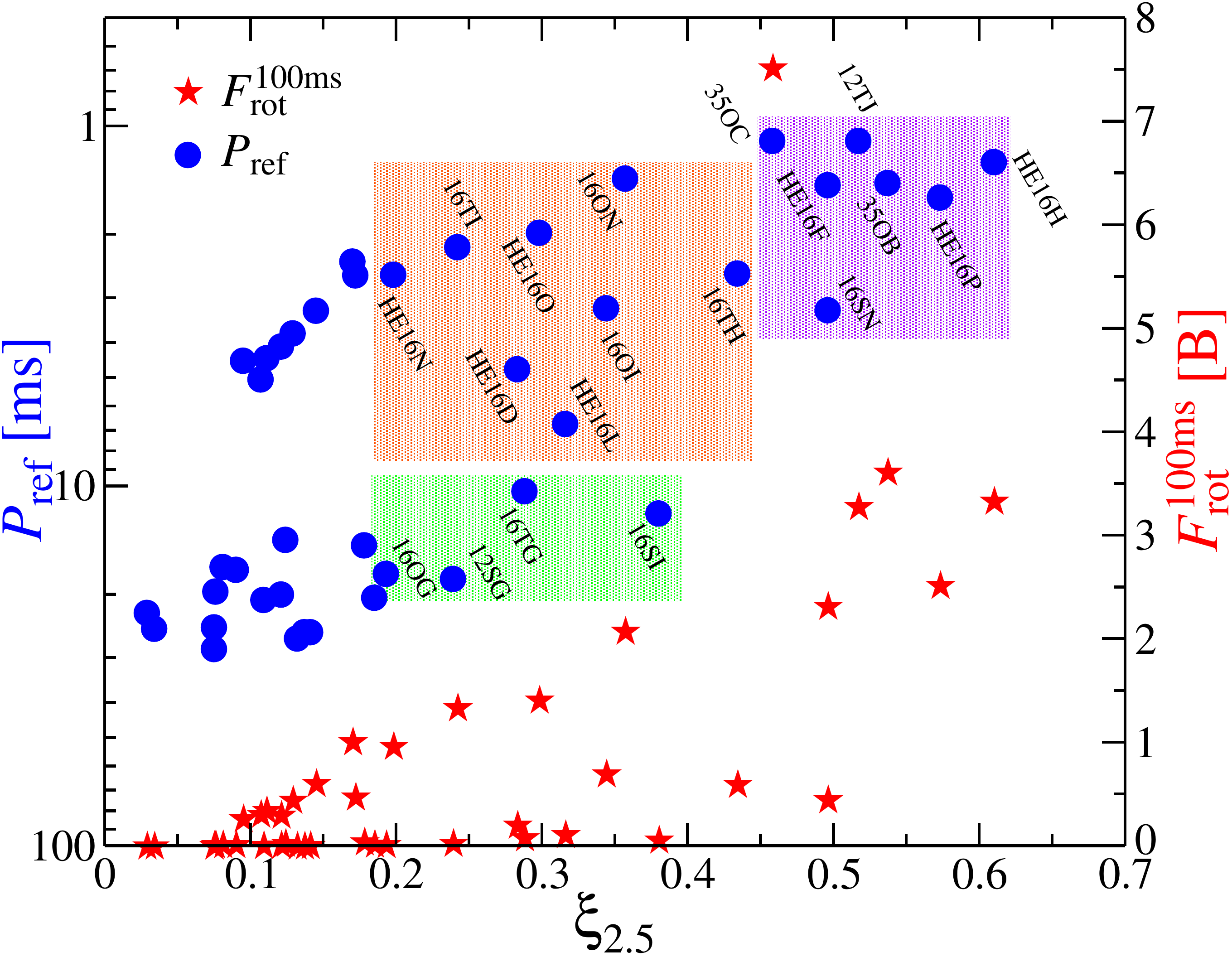}
\caption{Reference \pns\ spin period, $P_{\rm ref}$, taken at the
  onset of explosion (left axis, blue dots; Eq.~\ref{eq:omegabar}) and
  the free energy stored in differential rotation 100\,ms after bounce
  $F_\mathrm{rot}^\mathrm{100ms}$ (right axis, red stars;
  Eq.~\ref{eq:freeenergy}) versus bounce compactness $\xi_{2.5}$ for
  all rotating models in Table~\ref{tab:results}.  While models with a
  low bounce compactness show a diversity in core-rotation properties,
  those with a high bounce compactness systematically have short spin
  periods and a large budget of free energy stored in the differential
  rotation. Shaded boxes refer to specific groupings of
    models discussed in the text. Using $\xi_{2.5} > 0.45$ as a
  black-hole formation criterion for non-rotating progenitors, we can
  qualitatively compare the reference spin periods of this figure to
  \citet{metzger:11}, who sketches the outcome of collapse as a
  function of progenitor spin and mass. From this, one would predict
  that none of the LGRB progenitor models studied here formed black
  holes.}
  \label{fig:freeenergy_period}
\end{figure}

Additionally, we can discuss the potential for these models to form a
LGRB via the proto-magnetar model. Using
Eqs.~\ref{eq:freeenergy}~-~\ref{eq:omegabar}, we calculate the free
energy available in differential rotation at 100\,ms after bounce.  We
also calculate a reference spin period ($P_\text{ref}
=2\pi/\bar{\omega}$), measured at the onset of the neutrino driven
explosion, by assuming solid-body rotation for the entire \pns\ with
the same total angular momentum and moment of inertia. The
corresponding values are given in Table~\ref{tab:results}.  In
Fig.~\ref{fig:freeenergy_period}, we show the free energy available in
differential rotation at 100\,ms and the spin period of the \pns\ at
the onset of explosion. The total rotational energy of the \pns,
estimated as $I\bar{\omega}^2/2$, will increase as the \pns\ cools and
contracts. In models 12SG, 16OG, 16SI, and 16TG, which
  are contained within the green (lightest shade) box of
  Fig.~\ref{fig:freeenergy_period}, $\lesssim\,$0.1B of free energy
could be extracted from differential rotation via the MRI and
converted to explosion energy, much less than is needed for a
magneto-rotational explosion.  Also, the \pns\ spin periods are
$\gtrsim$\,10\,ms,\footnote{Even taking into account the spin up due
  to the PNS cooling and contraction, which will decrease the moment
  of inertia from the value in Table~\ref{tab:results} to $\sim 0.4
  M_\text{PNS} R_\text{PNS}^2 \sim
  1.6\times10^{45}(M/1.4\,M_\odot)(R/12\,\text{km})^2$
  \citep{metzger:11}, or roughly a factor of 2, the spin periods are
  $\gtrsim 5\,$ms.}  significantly larger than the $\lesssim 2\,$ms
periods required for the proto-magnetar model to reproduce classical
LGRB energies \citep{metzger:11}.

The second category are models with a larger angular-momentum budget
but unfavorable bounce compactness. Although compact enough to lead to
black-hole formation within 3.5\,s of core bounce, we find that the
predicted critical heating efficiencies are similar to that expected
for a standard 15\,\msun\ non-rotating RSG progenitor star
\citep{oconnor:11}.  These properties make them unlikely collapsar
progenitors, but in contrast, make them ideal candidates for
proto-magnetar formation, and perhaps LGRBs through that channel.
These models include 16OI ($\xi_{2.5} = 0.344$), 16ON ($\xi_{2.5} =
0.357$), 16TH ($\xi_{2.5} = 0.434$), 16TI ($\xi_{2.5} = 0.242$), HE16D
($\xi_{2.5} = 0.283$), HE16L ($\xi_{2.5} = 0.316$), HE16N ($\xi_{2.5}
= 0.198$), and HE16O ($\xi_{2.5} = 0.298$) and are
  contained in the orange (medium shade) box of
  Fig.~\ref{fig:freeenergy_period}. In addition to having critical
heating efficiencies similar to what is needed to explode typical
low-mass massive stars, the free energy available in rotation is
$\mathcal{O}$(1\,B). This energy may be converted to explosion energy
via the magneto-rotational mechanism.  The spin period of these \pns s
is in the range 1-6\,ms, thus on the order of what is needed for the
proto-magnetar model of LGRBs \citep{metzger:11}.

Eventually, the fastest rotating progenitor models evolved with a
strongly inhibited stellar-wind mass loss represent more suitable
collapsar candidates, although each model has caveats.  This set is
contained in the purple (darkest shade) box of
Fig.~\ref{fig:freeenergy_period} and includes models 12TJ ($\xi_{2.5}
= 0.517$), 16SN ($\xi_{2.5} = 0.496$), 35OB ($\xi_{2.5} = 0.537$), and
35OC ($\xi_{2.5} = 0.458$).  Model 12TJ will form a
2.37\,\msun\ (gravitational mass) \bh\ 0.85\,s after core bounce,
followed by a Keplerian disk after 2.64\,s, with a potential ejecta
mass of 8.57\,\msun. However, much like the models in the previous
category, model 12TJ has $\sim$3\,B of free energy available in
rotation that may lead to a magneto-rotational explosion early-on,
preventing collapsar formation.  This model is evolved at 1\% solar
metallicity, with an additional mass-loss rate scaling of 0.1,
equivalent to an overall evolution at $10^{-4}$ solar metallicity,
much below that observed for LGRB/SN sites.  We find that models
HE16F, HE16H, and HE16P have similar characteristics to model 12TJ.
Model 16SN forms a 2.27\,\msun\ \bh\ 0.78\,s after bounce.  Being
evolved at an effective metallicity of 0.01 solar, it has a lower
angular-momentum budget at death and is thus more likely to avoid a
magneto-rotational explosion. However, it forms a Keplerian disk only
40.6\,s after core bounce, with only 1.77\,\msun\ left over for the SN
ejecta. Such characteristics might in fact be more amenable to
reproduce recent observations of LGRB/SNe characterized by a very
early and narrow light-curve peak, as witnessed for example for
GRB100316D/SN 2010bh \citep{chornock:10}.  They may even explain why
no SN is found in association with some nearby LGRBs \citep{fynbo:06}.
Finally, models 35OB and 35OC form a black hole within 0.78 and
0.97\,s of bounce, respectively.  Model 35OB will accrete
$\sim$\,16.4\,\msun\ before a Keplerian disk forms $\sim$31.5\,s after
the onset of collapse, 4.8\,\msun\ is then available for the SN
ejecta.  With the 35OC model, a disk forms very quickly after
collapse, in 4.8\,s, and a significant amount of mass is exterior to
the disk, $23.6M_\odot$, and thus much too large to accommodate
inferred LGRB/SN ejecta masses. However, the propensity to collapsar
formation of the 35OB and 35OC model may be ill-founded if the MRI is
successful at powering a magneto-rotational explosion. The free energy
available in rotation is huge, i.e.\ on the order of 4-7.5\,B.  In
fact, in the 2D magneto-hydrodynamic simulations of
\citet{dessart:08a} based on the 35OC model, it was found that,
despite the large progenitor compactness, a magneto-rotational
explosion was initiated $\sim$200\,ms\ after core bounce and that the
\pns\ mass decreased thereafter, never reaching the mass threshold for
black-hole formation. In our models, the protoneutron stars in models
35OB and 35OC have $\sim 30-70\,$B of total rotational energy at the
onset of explosion, amply matching the inferred energies of observed
hypernovae.  These inferences are based on the assumption that energy
extraction from the differentially-rotating layers at the
proto-neutron star surface is efficient and can power an
explosion. Failing to do so, black-hole formation would result,
although the question of energy extraction from the disk for the
powering of a GRB would then arise.  Detailed multi-dimensional
core-collapse simulations need to be carried out to investigate the
efficiencies of magnetic/rotational/hydrodynamical instabilities for
the transport of angular momentum and the extraction of rotational
energy.

\section{Discussion}
\label{sect_conc}

In this paper, we have performed 1D general-relativistic
hydrodynamical simulations with \code{GR1D} of the collapse, bounce,
and post-bounce phases of the LGRB candidates of \citet{woosley:06} to
investigate their propensity to black-hole and disk formation.  We find that
these progenitors are at odds with the proposed criteria for a
collapsar progenitor or with the inferred properties of observed
LGRB/SNe, namely a H-deficient He-poor WR star with a massive iron
core (equivalent to a large compactness), a large angular momentum to
form a Keplerian disk soon after black-hole formation, an ejecta of
$\sim$10\,\msun, and an evolution at about 0.1 solar metallicity.  A
critical aspect that we focus on in this study is the compactness of
the progenitor cores at bounce, a quantity that helps diagnose the
likelihood of black-hole formation.

We group the \citet{woosley:06} models in different categories
according to their suitability for producing collapsars:

\begin{enumerate}
\item Models with a dimensionless Kerr spin parameter greater
  than unity at an enclosed mass of 3\,\msun, i.e. the models
  identified by \citet{woosley:06} as having the best potential for
  collapsar formation, fail to collapse when evolved with \code{GR1D}.
  Their cores are so fast spinning that the associated centrifugal
  acceleration leads them into expansion. All these models are evolved
  until death in \code{KEPLER} without magnetic fields and centrifugal
  forces, which seems questionable given the unrealistically short
  spin periods at collapse.
\item Models evolved with magnetic fields produce much lower
  rotation rates and most collapse with \code{GR1D}.
\item Of those evolved with magnetic fields, models with moderate
  rotation produce progenitors with a small compactness comparable to
  that characterizing the low-mass massive-star models proposed as
  progenitors of garden-variety core-collapse SNe. A small fraction of
  these is endowed with sufficient angular momentum to make a
  proto-magnetar, and thus a potential channel for producing LGRBs.
\item A few models (12TJ, 16SN) with the fastest rotation possess a
  large compactness favorable for black-hole formation and sufficient
  angular momentum for the formation of a Keplerian disk, but they
  require evolution at metallicities in the range 0.0001-0.01\,Z$_{\rm
    sol}$, significantly lower than the metallicity of a few tenths
  solar or even higher at which these LGRB/SNe are found
  \citep{modjaz:08,levesque:10a,levesque:10b}.  Although in many
  respects very attractive for forming a collapsar (if we ignore its
  huge core angular momentum), model 35OC is characterized by a large
  envelope mass of $\sim$23\,\msun, which is a factor 2-10 times
  larger than the inferred ejecta mass of LGRB/SNe discovered so far
  (for a summary, see \citealt{berger:11}).  We note that these models
  have a large angular momentum in the core, as models in the previous
  category, and may thus experience a magneto-rotational explosion
  preventing collapsar formation.
\end{enumerate}

Our quantitative study thus spells out the various
shortcomings of these progenitor stars for producing collapsars. Even
in those models that have the right compactness for black-hole formation and
sufficient angular momentum for disk formation, it is still unresolved
today how they would avoid the magneto-rotational mechanism of
explosion that is used to explain hypernovae \citep{leblanc:70, bisno:76,
wheeler:00, yamada:04, moiseenko:06,burrows:07b, dessart:08a, takiwaki:11}
The difficulty of forming a \bh\ and avoiding a magneto-rotational explosion
in fast-rotating cores, at least in the models of \cite{woosley:06}, lends credence
to the proto-magnetar model of LGRB/SNe.

Uncertainties in mass loss at low metallicity, and in particular
during transient phases of dynamical mass loss as observed in some
Luminous Blue Variable stars, is an issue, since it may completely
dominate the mass lost in the form of a weaker, but secular,
steady-state wind \citep{owocki:04}.  This uncertain mass-loss rate
plagues more severely the evolution of higher-mass stars, since
15-20\,\msun\ stars stay further away from the Eddington limit, and
overall lose little mass, even at solar metallicity.  By what
mechanism, at what rate, and during what phases a
100\,\msun\ star loses mass (and angular momentum) is much less well
known and this directly conditions the final mass and iron core mass
at collapse.

Overall, this suggests that studies of collapsar progenitors
would benefit from a second look. Angular momentum is key in the
current collapsar and proto-magnetar models, but there is a stiff
requirement on the progenitor compactness to speculate on its
propensity for forming a \bh, and thus for producing a LGRB through
one or the other channel.  A major step forward in resolving those
issues would be to conduct massive-star evolution with rotation,
centrifugal force, and magnetic fields always all the way to the
formation of a degenerate neutronized core on the verge of
collapse. This would allow a straightforward comparison of results
between groups, and an easy determination of the compactness using
\code{GR1D} to test the suitability of the core for black-hole
formation.

The ultimate check on the collapsar model requires multi-dimensional
simulations covering the whole evolution from progenitor collapse,
bounce, failed explosion during the \pns\ phase, formation of a
\bh\ followed by the formation of a Keplerian disk, and the powering
of a $\sim$\,10\,B SN ejecta. As we emphasize, black-hole
  formation is perhaps one of the most difficult steps in this
  sequence of events, and in that respect, renders the proto-magnetar
  channel quite attractive for the production of hypernovae and LGRBs.
  The diversity of LGRB/SNe, the existence of SN-less LGRBs and of
  LGRB-less hypernovae, may in fact call for a variety of formation
  channels for these rare events, including both collapsars and
  proto-magnetars.

\section*{Acknowledgements}
We acknowledge fruitful discussions with R.~Hirschi, A.~Beloborodov,
and T.~Piro. We also thank Stan Woosley for his comments on a draft
version of this paper. This research is supported in part by the
National Science Foundation under grand Nos.\ AST-0855535 and
OCI-0905046 and by the Sherman Fairchild Foundation.  E.O. is
supported in part by a post-graduate fellowship from the Natural
Sciences and Engineering Research Council of Canada (NSERC).  The
computations were performed at Caltech's Center for Advanced Computing
Research on the cluster ``Zwicky'' funded through NSF grant
no.\ PHY-0960291 and the Sherman Fairchild Foundation.


\begin{thebibliography}{92}
\expandafter\ifx\csname natexlab\endcsname\relax\def\natexlab#1{#1}\fi

\bibitem[{{Akiyama} {et~al.}(2003){Akiyama}, {Wheeler}, {Meier}, \&
  {Lichtenstadt}}]{akiyama:03}
{Akiyama}, S., {Wheeler}, J.~C., {Meier}, D.~L., \& {Lichtenstadt}, I. 2003,
  \apj, 584, 954

\bibitem[{{Aloy} {et~al.}(2000){Aloy}, {M{\"u}ller}, {Ib{\'a}{\~n}ez},
  {Mart{\'{\i}}}, \& {MacFadyen}}]{aloy:00}
{Aloy}, M.~A., {M{\"u}ller}, E., {Ib{\'a}{\~n}ez}, J.~M., {Mart{\'{\i}}},
  J.~M., \& {MacFadyen}, A. 2000, \apjl, 531, L119

\bibitem[{{Balbus} \& {Hawley}(1991)}]{balbus:91}
{Balbus}, S.~A., \& {Hawley}, J.~F. 1991, \apj, 376, 214

\bibitem[{{Bardeen} {et~al.}(1972){Bardeen}, {Press}, \&
  {Teukolsky}}]{bardeen:72}
{Bardeen}, J.~M., {Press}, W.~H., \& {Teukolsky}, S.~A. 1972, \apj, 178, 347

\bibitem[{{Berger} {et~al.}(2011){Berger}, {Chornock}, {Holmes}, {Foley},
  {Cucchiara}, {Wolf}, {Podsiadlowski}, {Fox}, \& {Roth}}]{berger:11}
{Berger}, E., {Chornock}, R., {Holmes}, T.~R., {et~al.} 2011, \apj, 743, 204

\bibitem[{{Bethe}(1990)}]{bethe:90}
{Bethe}, H.~A. 1990, Rev. Mod. Phys., 62, 801

\bibitem[{{Bethe} \& {Wilson}(1985)}]{bethewilson:85}
{Bethe}, H.~A., \& {Wilson}, J.~R. 1985, \apj, 295, 14

\bibitem[{{Bisnovatyi-Kogan} {et~al.}(1976){Bisnovatyi-Kogan}, {Popov}, \&
  {Samokhin}}]{bisno:76}
{Bisnovatyi-Kogan}, G.~S., {Popov}, I.~P., \& {Samokhin}, A.~A. 1976, \apss,
  41, 287

\bibitem[{{Bucciantini} {et~al.}(2007){Bucciantini}, {Quataert}, {Arons},
  {Metzger}, \& {Thompson}}]{bucciantini:07}
{Bucciantini}, N., {Quataert}, E., {Arons}, J., {Metzger}, B.~D., \&
  {Thompson}, T.~A. 2007, \mnras, 380, 1541

\bibitem[{{Bucciantini} {et~al.}(2008){Bucciantini}, {Quataert}, {Arons},
  {Metzger}, \& {Thompson}}]{bucciantini:08}
---. 2008, \mnras, 383, L25

\bibitem[{{Bucciantini} {et~al.}(2009){Bucciantini}, {Quataert}, {Metzger},
  {Thompson}, {Arons}, \& {Del Zanna}}]{bucciantini:09}
{Bucciantini}, N., {Quataert}, E., {Metzger}, B.~D., {et~al.} 2009, \mnras,
  396, 2038

\bibitem[{{Buras} {et~al.}(2006{\natexlab{a}}){Buras}, {Janka}, {Rampp}, \&
  {Kifonidis}}]{buras:06b}
{Buras}, R., {Janka}, H.-T., {Rampp}, M., \& {Kifonidis}, K.
  2006{\natexlab{a}}, \aap, 457, 281

\bibitem[{{Buras} {et~al.}(2006{\natexlab{b}}){Buras}, {Rampp}, {Janka}, \&
  {Kifonidis}}]{buras:06a}
{Buras}, R., {Rampp}, M., {Janka}, H.-T., \& {Kifonidis}, K.
  2006{\natexlab{b}}, Astron. Astrophys., 447, 1049

\bibitem[{{Burrows}(1986)}]{burrows:86bh}
{Burrows}, A. 1986, \apj, 300, 488

\bibitem[{{Burrows} {et~al.}(2007){Burrows}, {Dessart}, {Livne}, {Ott}, \&
  {Murphy}}]{burrows:07b}
{Burrows}, A., {Dessart}, L., {Livne}, E., {Ott}, C.~D., \& {Murphy}, J. 2007,
  \apj, 664, 416

\bibitem[{{Burrows} {et~al.}(1995){Burrows}, {Hayes}, \& {Fryxell}}]{bhf:95}
{Burrows}, A., {Hayes}, J., \& {Fryxell}, B.~A. 1995, \apj, 450, 830

\bibitem[{{Burrows} \& {Lattimer}(1983)}]{burrows:83}
{Burrows}, A., \& {Lattimer}, J.~M. 1983, \apj, 270, 735

\bibitem[{{Burrows} {et~al.}(2006){Burrows}, {Livne}, {Dessart}, {Ott}, \&
  {Murphy}}]{burrows:06}
{Burrows}, A., {Livne}, E., {Dessart}, L., {Ott}, C.~D., \& {Murphy}, J. 2006,
  Astrophys. J., 640, 878

\bibitem[{{Cantiello} {et~al.}(2007){Cantiello}, {Yoon}, {Langer}, \&
  {Livio}}]{cantiello:07}
{Cantiello}, M., {Yoon}, S., {Langer}, N., \& {Livio}, M. 2007, \aap, 465, L29

\bibitem[{{Chornock} {et~al.}(2010){Chornock}, {Berger}, {Levesque},
  {Soderberg}, {Foley}, {Fox}, {Frebel}, {Simon}, {Bochanski}, {Challis},
  {Kirshner}, {Podsiadlowski}, {Roth}, {Rutledge}, {Schmidt}, {Sheppard}, \&
  {Simcoe}}]{chornock:10}
{Chornock}, R., {Berger}, E., {Levesque}, E.~M., {et~al.} 2010, ArXiv:1004.2262

\bibitem[{{Demorest} {et~al.}(2010){Demorest}, {Pennucci}, {Ransom}, {Roberts},
  \& {Hessels}}]{demorest:10}
{Demorest}, P.~B., {Pennucci}, T., {Ransom}, S.~M., {Roberts}, M.~S.~E., \&
  {Hessels}, J.~W.~T. 2010, \nat, 467, 1081

\bibitem[{{Dessart} {et~al.}(2008){Dessart}, {Burrows}, {Livne}, \&
  {Ott}}]{dessart:08a}
{Dessart}, L., {Burrows}, A., {Livne}, E., \& {Ott}, C.~D. 2008, \apjl, 673,
  L43

\bibitem[{{Dessart} \& {Hillier}(2008)}]{dessart:08}
{Dessart}, L., \& {Hillier}, D.~J. 2008, \mnras, 383, 57

\bibitem[{{Dessart} {et~al.}(2012){Dessart}, {Hillier}, {Li}, \&
  {Woosley}}]{dessart:12}
{Dessart}, L., {Hillier}, D.~J., {Li}, C., \& {Woosley}, S. 2012, submitted to
  MNRAS

\bibitem[{{Dimmelmeier} {et~al.}(2008){Dimmelmeier}, {Ott}, {Marek}, \&
  {Janka}}]{dimmelmeier:08}
{Dimmelmeier}, H., {Ott}, C.~D., {Marek}, A., \& {Janka}, H.-T. 2008, \prd, 78,
  064056

\bibitem[{{Fischer} {et~al.}(2009){Fischer}, {Whitehouse}, {Mezzacappa},
  {Thielemann}, \& {Liebend{\"o}rfer}}]{fischer:09a}
{Fischer}, T., {Whitehouse}, S.~C., {Mezzacappa}, A., {Thielemann}, F.-K., \&
  {Liebend{\"o}rfer}, M. 2009, \aap, 499, 1

\bibitem[{{Fynbo} {et~al.}(2006){Fynbo}, {Watson}, {Th{\"o}ne}, {Sollerman},
  {Bloom}, {Davis}, {Hjorth}, {Jakobsson}, {J{\o}rgensen}, {Graham},
  {Fruchter}, {Bersier}, {Kewley}, {Cassan}, {Castro Cer{\'o}n}, {Foley},
  {Gorosabel}, {Hinse}, {Horne}, {Jensen}, {Klose}, {Kocevski}, {Marquette},
  {Perley}, {Ramirez-Ruiz}, {Stritzinger}, {Vreeswijk}, {Wijers}, {Woller},
  {Xu}, \& {Zub}}]{fynbo:06}
{Fynbo}, J.~P.~U., {Watson}, D., {Th{\"o}ne}, C.~C., {et~al.} 2006, \nat, 444,
  1047

\bibitem[{{Georgy} {et~al.}(2009){Georgy}, {Meynet}, {Walder}, {Folini}, \&
  {Maeder}}]{georgy:09}
{Georgy}, C., {Meynet}, G., {Walder}, R., {Folini}, D., \& {Maeder}, A. 2009,
  \aap, 502, 611

\bibitem[{{Goldreich} \& {Weber}(1980)}]{goldreich:80}
{Goldreich}, P., \& {Weber}, S.~V. 1980, \apj, 238, 991

\bibitem[{{Guetta} \& {Della Valle}(2007)}]{guetta:07}
{Guetta}, D., \& {Della Valle}, M. 2007, \apjl, 657, L73

\bibitem[{{Hanke} {et~al.}(2011){Hanke}, {Marek}, {M\"uller}, \&
  {Janka}}]{hanke:11}
{Hanke}, F., {Marek}, A., {M\"uller}, B., \& {Janka}, H.-T. 2011, Submitted to
  the Astrophys. J., arXiv:1108.4355

\bibitem[{{Hebeler} {et~al.}(2010){Hebeler}, {Lattimer}, {Pethick}, \&
  {Schwenk}}]{hebeler:10}
{Hebeler}, K., {Lattimer}, J.~M., {Pethick}, C.~J., \& {Schwenk}, A. 2010,
  Phys. Rev. Lett., 105, 161102

\bibitem[{{Heger} {et~al.}(2000){Heger}, {Langer}, \& {Woosley}}]{heger:00}
{Heger}, A., {Langer}, N., \& {Woosley}, S.~E. 2000, \apj, 528, 368

\bibitem[{{Heger} {et~al.}(2005){Heger}, {Woosley}, \& {Spruit}}]{heger:05}
{Heger}, A., {Woosley}, S.~E., \& {Spruit}, H.~C. 2005, \apj, 626, 350

\bibitem[{{Herant} {et~al.}(1994){Herant}, {Benz}, {Hix}, {Fryer}, \&
  {Colgate}}]{herant:94}
{Herant}, M., {Benz}, W., {Hix}, W.~R., {Fryer}, C.~L., \& {Colgate}, S.~A.
  1994, \apj, 435, 339

\bibitem[{{Hirschi} {et~al.}(2004){Hirschi}, {Meynet}, \&
  {Maeder}}]{hirschi:04}
{Hirschi}, R., {Meynet}, G., \& {Maeder}, A. 2004, \aap, 425, 649

\bibitem[{{Hirschi} {et~al.}(2005){Hirschi}, {Meynet}, \&
  {Maeder}}]{hirschi:05}
---. 2005, \aap, 443, 581

\bibitem[{{Iwamoto} {et~al.}(1998){Iwamoto}, {Mazzali}, {Nomoto}, {Umeda},
  {Nakamura}, {Patat}, {Danziger}, {Young}, {Suzuki}, {Shigeyama},
  {Augusteijn}, {Doublier}, {Gonzalez}, {Boehnhardt}, {Brewer}, {Hainaut},
  {Lidman}, {Leibundgut}, {Cappellaro}, {Turatto}, {Galama}, {Vreeswijk},
  {Kouveliotou}, {van Paradijs}, {Pian}, {Palazzi}, \& {Frontera}}]{iwamoto:98}
{Iwamoto}, K., {Mazzali}, P.~A., {Nomoto}, K., {et~al.} 1998, \nat, 395, 672

\bibitem[{{Janka} \& {M\"uller}(1996)}]{jankamueller:96}
{Janka}, H.-T., \& {M\"uller}, E. 1996, \aap, 306, 167

\bibitem[{{Kitaura} {et~al.}(2006){Kitaura}, {Janka}, \&
  {Hillebrandt}}]{kitaura:06}
{Kitaura}, F.~S., {Janka}, H.-T., \& {Hillebrandt}, W. 2006, \aap, 450, 345

\bibitem[{{Komissarov} \& {Barkov}(2007)}]{komissarov:07}
{Komissarov}, S.~S., \& {Barkov}, M.~V. 2007, \mnras, 382, 1029

\bibitem[{Lattimer \& Swesty(1991)}]{lseos:91}
Lattimer, J.~M., \& Swesty, F.~D. 1991, {Nucl. Phys. A}, 535, 331

\bibitem[{{LeBlanc} \& {Wilson}(1970)}]{leblanc:70}
{LeBlanc}, J.~M., \& {Wilson}, J.~R. 1970, \apj, 161, 541

\bibitem[{{Levesque} {et~al.}(2010{\natexlab{a}}){Levesque}, {Kewley},
  {Berger}, \& {Zahid}}]{levesque:10b}
{Levesque}, E.~M., {Kewley}, L.~J., {Berger}, E., \& {Zahid}, H.~J.
  2010{\natexlab{a}}, \aj, 140, 1557

\bibitem[{{Levesque} {et~al.}(2010{\natexlab{b}}){Levesque}, {Soderberg},
  {Foley}, {Berger}, {Kewley}, {Chakraborti}, {Ray}, {Torres}, {Challis},
  {Kirshner}, {Barthelmy}, {Bietenholz}, {Chandra}, {Chaplin}, {Chevalier},
  {Chugai}, {Connaughton}, {Copete}, {Fox}, {Fransson}, {Grindlay}, {Hamuy},
  {Milne}, {Pignata}, {Stritzinger}, \& {Wieringa}}]{levesque:10a}
{Levesque}, E.~M., {Soderberg}, A.~M., {Foley}, R.~J., {et~al.}
  2010{\natexlab{b}}, \apjl, 709, L26

\bibitem[{{Liebend{\"o}rfer} {et~al.}(2004){Liebend{\"o}rfer}, {Messer},
  {Mezzacappa}, {Bruenn}, {Cardall}, \& {Thielemann}}]{liebendoerfer:04}
{Liebend{\"o}rfer}, M., {Messer}, O.~E.~B., {Mezzacappa}, A., {et~al.} 2004,
  \apjs, 150, 263

\bibitem[{{Limongi} \& {Chieffi}(2006)}]{limongi:06}
{Limongi}, M., \& {Chieffi}, A. 2006, \apj, 647, 483

\bibitem[{{Lindner} {et~al.}(2010){Lindner}, {Milosavljevi{\'c}}, {Couch}, \&
  {Kumar}}]{lindner:10}
{Lindner}, C.~C., {Milosavljevi{\'c}}, M., {Couch}, S.~M., \& {Kumar}, P. 2010,
  \apj, 713, 800

\bibitem[{{MacFadyen} \& {Woosley}(1999)}]{macfadyen:99}
{MacFadyen}, A.~I., \& {Woosley}, S.~E. 1999, \apj, 524, 262

\bibitem[{{MacFadyen} {et~al.}(2001){MacFadyen}, {Woosley}, \&
  {Heger}}]{macfadyen:01}
{MacFadyen}, A.~I., {Woosley}, S.~E., \& {Heger}, A. 2001, \apj, 550, 410

\bibitem[{{Maeder} \& {Meynet}(2000)}]{maeder:00}
{Maeder}, A., \& {Meynet}, G. 2000, \araa, 38, 143

\bibitem[{{Mazzali} {et~al.}(2002){Mazzali}, {Deng}, {Maeda}, {Nomoto},
  {Umeda}, {Hatano}, {Iwamoto}, {Yoshii}, {Kobayashi}, {Minezaki}, {Doi},
  {Enya}, {Tomita}, {Smartt}, {Kinugasa}, {Kawakita}, {Ayani}, {Kawabata},
  {Yamaoka}, {Qiu}, {Motohara}, {Gerardy}, {Fesen}, {Kawabata}, {Iye},
  {Kashikawa}, {Kosugi}, {Ohyama}, {Takada-Hidai}, {Zhao}, {Chornock},
  {Filippenko}, {Benetti}, \& {Turatto}}]{mazzali:02}
{Mazzali}, P.~A., {Deng}, J., {Maeda}, K., {et~al.} 2002, \apjl, 572, L61

\bibitem[{{Metzger}(2010)}]{metzger:10c}
{Metzger}, B.~D. 2010, in Astronomical Society of the Pacific Conference
  Series, Vol. 432, New Horizons in Astronomy: Frank N. Bash Symposium 2009,
  ed. {L.~M.~Stanford, J.~D.~Green, L.~Hao, \& Y.~Mao}, 81

\bibitem[{{Metzger} {et~al.}(2011){Metzger}, {Giannios}, {Thompson},
  {Bucciantini}, \& {Quataert}}]{metzger:11}
{Metzger}, B.~D., {Giannios}, D., {Thompson}, T.~A., {Bucciantini}, N., \&
  {Quataert}, E. 2011, \mnras, 413, 2031

\bibitem[{{Meynet} \& {Maeder}(2000)}]{meynet:00}
{Meynet}, G., \& {Maeder}, A. 2000, \aap, 361, 101

\bibitem[{{Meynet} \& {Maeder}(2005)}]{meynet:05}
---. 2005, \aap, 429, 581

\bibitem[{{Milosavljevi{\'c}} {et~al.}(2012){Milosavljevi{\'c}}, {Lindner},
  {Shen}, \& {Kumar}}]{milosavljevic:12}
{Milosavljevi{\'c}}, M., {Lindner}, C.~C., {Shen}, R., \& {Kumar}, P. 2012,
  \apj, 744, 103

\bibitem[{{Modjaz} {et~al.}(2008){Modjaz}, {Kewley}, {Kirshner}, {Stanek},
  {Challis}, {Garnavich}, {Greene}, {Kelly}, \& {Prieto}}]{modjaz:08}
{Modjaz}, M., {Kewley}, L., {Kirshner}, R.~P., {et~al.} 2008, \aj, 135, 1136

\bibitem[{{Moiseenko} {et~al.}(2006){Moiseenko}, {Bisnovatyi-Kogan}, \&
  {Ardeljan}}]{moiseenko:06}
{Moiseenko}, S.~G., {Bisnovatyi-Kogan}, G.~S., \& {Ardeljan}, N.~V. 2006,
  \mnras, 370, 501

\bibitem[{{Murphy} \& {Burrows}(2008)}]{murphy:08}
{Murphy}, J.~W., \& {Burrows}, A. 2008, \apj, 688, 1159

\bibitem[{{Nordhaus} {et~al.}(2010){Nordhaus}, {Burrows}, {Almgren}, \&
  {Bell}}]{nordhaus:10}
{Nordhaus}, J., {Burrows}, A., {Almgren}, A., \& {Bell}, J. 2010, \apj, 720,
  694

\bibitem[{{Obergaulinger} {et~al.}(2009){Obergaulinger}, {Cerd{\'a}-Dur{\'a}n},
  {M{\"u}ller}, \& {Aloy}}]{obergaulinger:09}
{Obergaulinger}, M., {Cerd{\'a}-Dur{\'a}n}, P., {M{\"u}ller}, E., \& {Aloy},
  M.~A. 2009, \aap, 498, 241

\bibitem[{{O'Connor} \& {Ott}(2010)}]{oconnor:10}
{O'Connor}, E., \& {Ott}, C.~D. 2010, \cqg, 27, 114103

\bibitem[{{O'Connor} \& {Ott}(2011)}]{oconnor:11}
---. 2011, \apj, 730, 70

\bibitem[{{Ott} {et~al.}(2006){Ott}, {Burrows}, {Thompson}, {Livne}, \&
  {Walder}}]{ott:06spin}
{Ott}, C.~D., {Burrows}, A., {Thompson}, T.~A., {Livne}, E., \& {Walder}, R.
  2006, Astrophys. J. Suppl. Ser., 164, 130

\bibitem[{{Ott} {et~al.}(2011){Ott}, {Reisswig}, {Schnetter}, {O'Connor},
  {Sperhake}, {L{\"o}ffler}, {Diener}, {Abdikamalov}, {Hawke}, \&
  {Burrows}}]{ott:11a}
{Ott}, C.~D., {Reisswig}, C., {Schnetter}, E., {et~al.} 2011, Phys. Rev. Lett.,
  106, 161103

\bibitem[{{Owocki} {et~al.}(2004){Owocki}, {Gayley}, \& {Shaviv}}]{owocki:04}
{Owocki}, S.~P., {Gayley}, K.~G., \& {Shaviv}, N.~J. 2004, \apj, 616, 525

\bibitem[{{{\"O}zel} {et~al.}(2010){{\"O}zel}, {Baym}, \&
  {G{\"u}ver}}]{oezel:10b}
{{\"O}zel}, F., {Baym}, G., \& {G{\"u}ver}, T. 2010, \prd, 82, 101301

\bibitem[{{Pastorello} {et~al.}(2004){Pastorello}, {Zampieri}, {Turatto},
  {Cappellaro}, {Meikle}, {Benetti}, {Branch}, {Baron}, {Patat}, {Armstrong},
  {Altavilla}, {Salvo}, \& {Riello}}]{pastorello:04}
{Pastorello}, A., {Zampieri}, L., {Turatto}, M., {et~al.} 2004, \mnras, 347, 74

\bibitem[{{Pejcha} \& {Thompson}(2011)}]{pejcha:11}
{Pejcha}, O., \& {Thompson}, T.~A. 2011, Submitted to ApJ. arXiv:1103.4865

\bibitem[{{Petrovic} {et~al.}(2005){Petrovic}, {Langer}, {Yoon}, \&
  {Heger}}]{petrovic:05}
{Petrovic}, J., {Langer}, N., {Yoon}, S.-C., \& {Heger}, A. 2005, \aap, 435,
  247

\bibitem[{{Proga} {et~al.}(2003){Proga}, {MacFadyen}, {Armitage}, \&
  {Begelman}}]{proga:03}
{Proga}, D., {MacFadyen}, A.~I., {Armitage}, P.~J., \& {Begelman}, M.~C. 2003,
  \apjl, 599, L5

\bibitem[{{Sekiguchi} \& {Shibata}(2011)}]{sekiguchi:11a}
{Sekiguchi}, Y., \& {Shibata}, M. 2011, \apj, 737, 6

\bibitem[{{Spruit}(2002)}]{spruit:02}
{Spruit}, H.~C. 2002, \aap, 381, 923

\bibitem[{{Steiner} {et~al.}(2010){Steiner}, {Lattimer}, \&
  {Brown}}]{steiner:10}
{Steiner}, A.~W., {Lattimer}, J.~M., \& {Brown}, E.~F. 2010, \apj, 722, 33

\bibitem[{{Sumiyoshi} {et~al.}(2007){Sumiyoshi}, {Yamada}, \&
  {Suzuki}}]{sumiyoshi:07}
{Sumiyoshi}, K., {Yamada}, S., \& {Suzuki}, H. 2007, \apj, 667, 382

\bibitem[{{Takiwaki} \& {Kotake}(2011)}]{takiwaki:11}
{Takiwaki}, T., \& {Kotake}, K. 2011, \apj, 743, 30

\bibitem[{{Thompson} {et~al.}(2005){Thompson}, {Quataert}, \&
  {Burrows}}]{thompson:05}
{Thompson}, T.~A., {Quataert}, E., \& {Burrows}, A. 2005, \apj, 620, 861

\bibitem[{{Wellstein} \& {Langer}(1999)}]{wellstein:99}
{Wellstein}, S., \& {Langer}, N. 1999, \aap, 350, 148

\bibitem[{{Wheeler} {et~al.}(2000){Wheeler}, {Yi}, {H{\"o}flich}, \&
  {Wang}}]{wheeler:00}
{Wheeler}, J.~C., {Yi}, I., {H{\"o}flich}, P., \& {Wang}, L. 2000, \apj, 537,
  810

\bibitem[{{Woosley}(1993)}]{woosley:93}
{Woosley}, S.~E. 1993, \apj, 405, 273

\bibitem[{{Woosley}(2011)}]{woosley:11a}
---. 2011, ArXiv:1105.4193

\bibitem[{{Woosley} {et~al.}(1999){Woosley}, {Eastman}, \&
  {Schmidt}}]{woosley:99}
{Woosley}, S.~E., {Eastman}, R.~G., \& {Schmidt}, B.~P. 1999, \apj, 516, 788

\bibitem[{{Woosley} \& {Heger}(2006)}]{woosley:06}
{Woosley}, S.~E., \& {Heger}, A. 2006, Astrophys. J., 637, 914

\bibitem[{{Woosley} \& {Heger}(2007)}]{woosley:07}
---. 2007, \physrep, 442, 269

\bibitem[{{Yamada} \& {Sawai}(2004)}]{yamada:04}
{Yamada}, S., \& {Sawai}, H. 2004, \apj, 608, 907

\bibitem[{{Yoon} \& {Langer}(2005)}]{yoon:05}
{Yoon}, S.-C., \& {Langer}, N. 2005, \aap, 443, 643

\bibitem[{{Yoon} {et~al.}(2006){Yoon}, {Langer}, \& {Norman}}]{yoon:06}
{Yoon}, S.-C., {Langer}, N., \& {Norman}, C. 2006, \aap, 460, 199

\bibitem[{{Zahn} {et~al.}(2007){Zahn}, {Brun}, \& {Mathis}}]{zahn:07}
{Zahn}, J.-P., {Brun}, A.~S., \& {Mathis}, S. 2007, \aap, 474, 145

\bibitem[{{Zhang} {et~al.}(2004){Zhang}, {Woosley}, \& {Heger}}]{zhang:04}
{Zhang}, W., {Woosley}, S.~E., \& {Heger}, A. 2004, \apj, 608, 365

\end{thebibliography}
\end{document}